\documentclass[12pt]{article}
\usepackage{amsmath,amsthm}
\usepackage{graphicx}
\usepackage{subfigure}
\usepackage{amssymb,latexsym}

\DeclareGraphicsExtensions{.pdf,.png,.jpg}

\newtheorem{theorem}{Theorem}[section]
\newtheorem{lemma}[theorem]{Lemma}

\newtheorem{definition}[theorem]{Definition}
\theoremstyle{remark}
\newtheorem{remark}[theorem]{Remark}

\newcommand{\bi}{\begin{itemize}}
\newcommand{\ei}{\end{itemize}}
\newcommand{\bt}{\begin{theorem}}
\newcommand{\et}{\end{theorem}}
\newcommand{\bp}{\begin{proof}}
\newcommand{\ep}{\end{proof}}
\newcommand{\be}{\begin{equation}}
\newcommand{\ee}{\end{equation}}
\newcommand{\ben}{\begin{enumerate}}
\newcommand{\een}{\end{enumerate}}

\newcommand{\Cbar}{\mathbb{\overline{C}}}

\newcommand{\N}{\mathbb N}

\newcommand{\e}{\varepsilon}

\renewcommand{\O}{\Omega}

\renewcommand{\b}{\beta}

\renewcommand{\a}{\alpha}

\newcommand{\g}{\gamma}

\renewcommand{\and}{\text{~~ and ~~}}
\renewcommand{\part}{\partial}

\newcommand{\sech}{{\rm sech~}}

\newcommand{\eps}{\varepsilon}
\newcommand{\mm}{{\mu\over 2}}

\renewcommand{\gg}{\gamma}
\newcommand{\ggh}{\hat\gamma}

\newcommand{\bb}{\vec\beta}
\newcommand{\vb}{\vec\beta}
\newcommand{\vDb}{\Delta\vec\beta}
\newcommand{\va}{\vec\alpha}

\begin{document}

\title{Perturbation of Riemann-Hilbert jump contours: smooth parametric dependence with application to semiclassical focusing NLS}

\author{Sergey Belov\footnote{ Department of Mathematics, Rice University, Houston, TX 77005, e-mail: belov@rice.edu},
Stephanos Venakides \footnote{ Department of Mathematics, Duke University, Durham, NC 27708, e-mail: ven@math.duke.edu.
SV thanks the NSF for supporting this research under grant
DMS-0707488. Part of the research was conducted while SV held a
Catedra de Excelencia at the Universidad Carlos III de Madrid and
while he was a visitor of the Archimedes Center for Modeling, Analysis
and Computation (ACMAC) at the University of Crete. SV thanks both
institutions.}}

\maketitle

\begin{abstract}
A perturbation of a class of scalar Riemann-Hilbert problems (RHPs)
with the jump contour as a finite union of oriented simple arcs in the complex plane
and the jump function with a $z\log z$ type singularity on the jump contour is considered.
The jump function and the jump contour are assumed to depend on a vector of external
parameters $\vec\beta$. We prove that if the RHP has a solution at some value $\vec\beta_0$
then the solution of the RHP is uniquely defined in a some neighborhood of $\vec\beta_0$
and is smooth in $\vec\beta$. This result is applied to the case of semiclassical focusing
NLS.

\end{abstract}

\section{Introduction}
We study a type of scalar Riemann-Hilbert problem that appears in
semiclassical or small dispersion calculations of integrable
systems. We prove the smooth dependence of the solution on a crucial
parameter. Our direct motivation comes from
 the semiclassical focusing NLS equation
\footnote{the equation as written here corrects a mistake in the
numerical coefficients of equation (1.1) of \cite{TVZzero}.}
\cite{CT,Lyng,KMM,MillerKamvissis}
\begin{equation}
i\eps\partial_t q+\eps^2\partial_x^2 q +2|q|^2q=0
\label{eq:1.1_NLS}
\end{equation}
with the initial condition

\begin{equation}
q(x,0)=A(x)e^{\frac{i\mu}{\eps}S(x)},\ \ \ A(x)>0,\ \ \ \mu\ge 0  \label{IC_NLS}
\end{equation}
in the limit as $\eps\to 0$.
For any value of $\e>0$, the solution process requires solving
a $2\times 2$ matrix Riemann-Hilbert problem (RHP) in the complex plane of the
spectral parameter $z$ of an underlying Lax pair operator \cite{TVZzero, TVZlong2}.
The quantities $x$, $t$, and $\mu$ appear as parameters. The Riemann-Hilbert approach
is relevant to other integrable systems in general as established by \cite{Shabat} and
it is a major tool in the asymptotic analysis of integrable systems as
established by the discovery of the steepest descent method \cite{DZ1,DZ2}. The asymptotic
methods also apply to orthogonal polynomial asymptotics \cite{Deift_book, DKMVZ_strong},
and through \cite{FIK1,FIK2}, to random matrices \cite{BDJ, DKMVZ_uniform, DJ, EM}.

When $\eps\to 0$, the steepest descent method \cite{DZ1,DZ2},
with the implementation of the $g$-function mechanism \cite{DVZ1,TVZzero},
reduces the asymptotic calculation to a $2\times 2$ "model" RHP which is exactly
solvable. The reduction occurs through a series of transformations
of the original RHP. These are facilitated by factorizations of the RHP jump-matrix
and appropriate RHP contour deformations. As a result,
the main contributions to the solution are isolated. Higher order
contributions can be calculated iteratively. Characteristic
of nonlinearity, the main contributions are linked to \emph{arcs} in the complex plane
that form the "contributing" part of the
contour of the model RHP. The process is in the spirit of the steepest
descent method for the asymptotic evaluation of integrals, where, however,
the contributions are localized near points.

In the nonlinear problem,
the arc endpoints are crucial quantities
that control the shape of the nonlinear waveforms that develop in the small $O(\eps)$
spatiotemporal scale. The number and the position of these endpoints in the
complex plane vary  with $x$ and $t$ in the large $O(1)$
scale, thus, modulating the waveform in space-time. They are
often referred to as the modulation parameters or as branchpoints, as
they arise naturally as the branchpoints of a square root.

The $g$-function mechanism reduces this process to
solving a scalar RHP and, thus, determining the branchpoints and
all other data needed for the model problem. Conceptually,
the scalar RHP problem  identifies how crucial features
of  the eigenfunctions and of the potential vary in the large
space-time scale
and facilitates the rigorous derivation of the solution of the
matrix RHP. It is, thus, an analogous entity to the
eikonal equation of linear PDEs. The function $g$ that
solves the scalar RHP problem is, essentially, the same phase
function that appears in the direct semiclassical scattering problem.
In the RHP it is  considered as a function of the spectral variable,
at fixed $x$ and $t$. The eikonal equation it addresses it
as a function of $x$ and $t$ at a fixed value of the spectral variable.

The smooth perturbation of the solution of the scalar RHP
with respect to parameters $x$ and $t$ was obtained in \cite{TV_det}, excluding a crucial parameter $\mu$ at which
the input function $f(z)$ to the scalar RHP had a $z\log z$
singularity. The following formula for the branch points (see below for the
definition of $\alpha_j$ and other quantities in the formula) was derived
\begin{equation}
\frac{\partial\alpha_j}{\partial\beta_k}(\vb)=-\frac{2\pi i \frac{\partial K(\alpha_j,\vec\alpha,\bb)}{\partial\beta_k}}{D(\vec\alpha,\bb) \oint_{\ggh(\beta_1)}\frac{f'(\zeta,\bb)}{(\zeta-\alpha_j(\bb))R(\zeta,\vec\alpha)}d\zeta}, \ \ \ \beta_2=x,\ \beta_3=t.
\end{equation}

We prove that the formula holds for the parameter $\beta_1=\mu$ as well and the
dependence is smooth, meaning that the contour, the jump matrix, and
the solution of the scalar RHP evolve smoothly. The significance of this
is that this and the continuation method will allow extending  long-time estimates
of the position of branchpoints to parameter regions beyond the
ones for which such estimates are known. In the NLS case this means
extending to regions of $\mu$ in which solitons appear.

The Perturbation theorem \ref{thm:perturb} has an immediate application to NLS
with inial data being a semiclassical approximation \cite{TVZzero} of
\begin{equation}
q(x,0)=-\sech(x)e^{-\frac{i\mu}{\eps}\int_0^x\tanh(s)ds},\ \ \ \mu\ge 0.  \label{our_IC_NLS}
\end{equation}
This family of initial conditions is interesting because of a transition
at $\mu=2$ with solitonless interval ($\mu\ge 2 $) and soliton interval
$0<\mu<2$. It has been studied in a number of papers \cite{TVZzero, TVZlong2, TVZ2}
The semiclassical limit has been completely analyzed for $\mu\ge 2$ while
for the soliton case $0<\mu<2$ the answer is known only for some finite times.
The RH approach runs into difficulties with the error estimates and
was not able to continue past a certain curve.
Numerical experiments have shown absence of any noticeable
transition in the behavior of end points $\alpha_j(\mu)$ at the critical
value $\mu=2$ \cite{BTV}. The Perturbation theorem for NLS \ref{thm:perturb_NLS} establishes
this fact rigorously.

\section{The scalar Riemann-Hilbert problem (RHP): background}

\begin{figure}
\begin{center}
{\includegraphics[height=6cm]{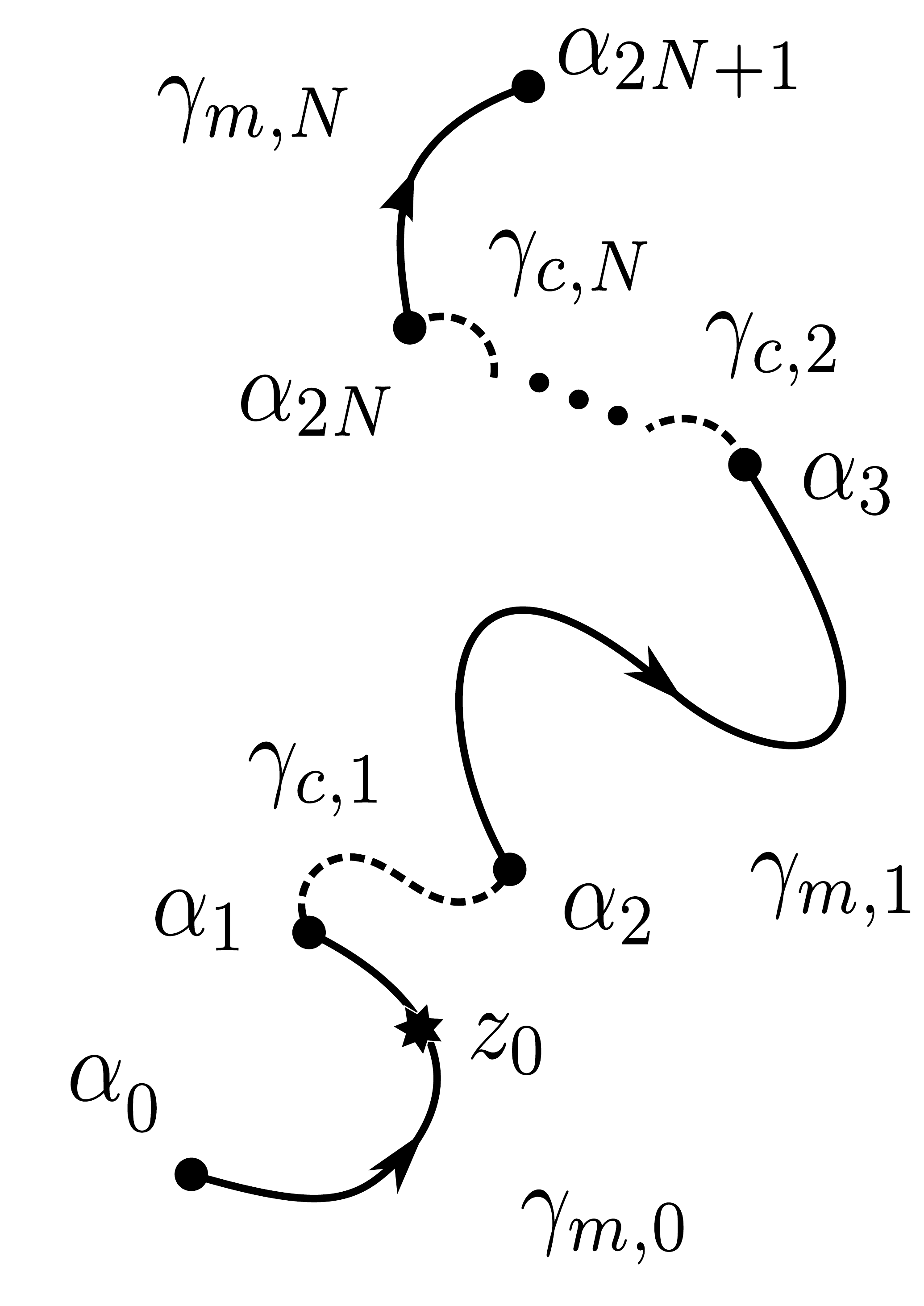}}
\caption{\label{fig:RHP_contours} RHP jump contours}
\end{center}
\end{figure}

The following scalar RHP is studied in \cite{TV_det}. In this section we provide an overview. The contour of RHP, labeled  $\g$ is
an unknown of the problem, except for the fact
 that it is a connected, finite, non self intersecting, typically open curve, that is partitioned  into $2N+1$
arcs $[\alpha_0,\alpha_1]$, $[\alpha_1,\alpha_2]$, $[\alpha_2,\alpha_3]$, $\ldots$   by the points $\a_0,\a_1,\ldots,
\a_{2N+1}$, that have the natural ordering of their indices $0 , 1 , 2 ,\ldots, 2N+1$ along the
oriented contour. The number $N=0, 1, 2,\ldots$ is given in the problem. A scalar function $g(z)$ is sought
that is analytic and bounded in $\Cbar\backslash\gamma$. The jump condition on
the $N+1$ {\it main arcs} $[\alpha_0,\alpha_1]$, $[\alpha_2,\alpha_3]$, $\ldots$, $[\alpha_{2N},\alpha_{2N+1}]$
(they include the arcs at the two ends of the contour) is
\begin{equation}
g_+ + g_- = f + W
\label{jumpmain}
\end{equation}
where $g_{\pm}(z)$ are the nontangential limiting values of $g(z)$ from the
positive/negative side of the contour, $f(z)$ is a given function, that constitutes the main input to the
problem, and $W$ is a real constant to be determined. The constant $W$
generally takes different values on different arcs.
The contour is sought in the domain of analyticity of $f(z)$,
except for  a \emph{finite} number of its points where $f(z)$ can
be non-analytic \cite{TV_det}. In the case of NLS, $f(z)$ must be determined
from the initial data (\ref{IC_NLS}).

The condition   on the
remaining  $N$  {\it complementary arcs}, that interlace with
the main arcs is
\begin{equation}
g_+ - g_- = \O
\label{jumpcomp}
\end{equation}
where $\O$ is again a real constant to be determined and takes different values on different arcs.

\begin{figure}
\begin{center}
\subfigure[Contour $\ggh$]{\includegraphics[height=6cm]{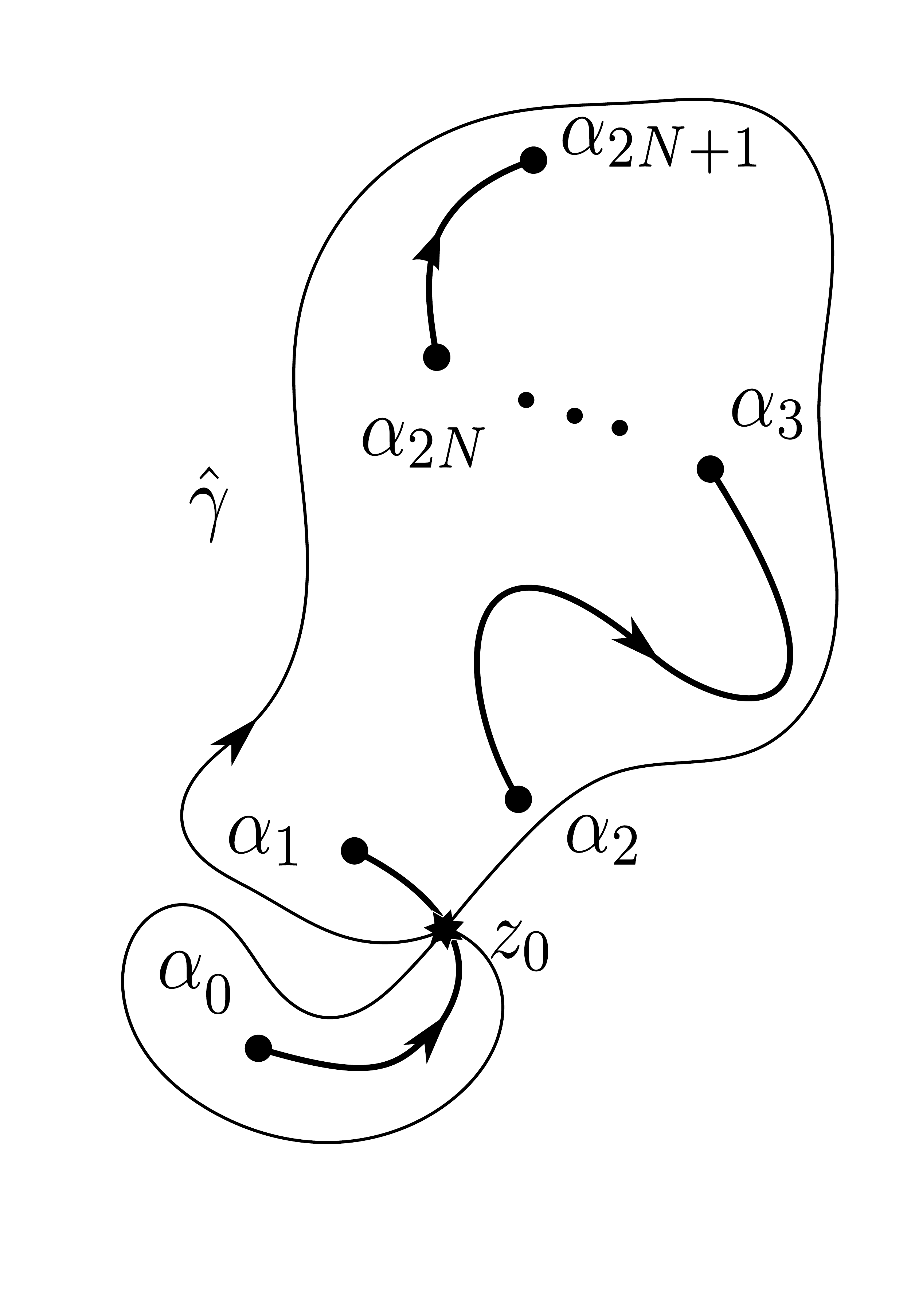}}
\subfigure[Contours $\ggh_{m,j}$ ]{\includegraphics[height=6cm]{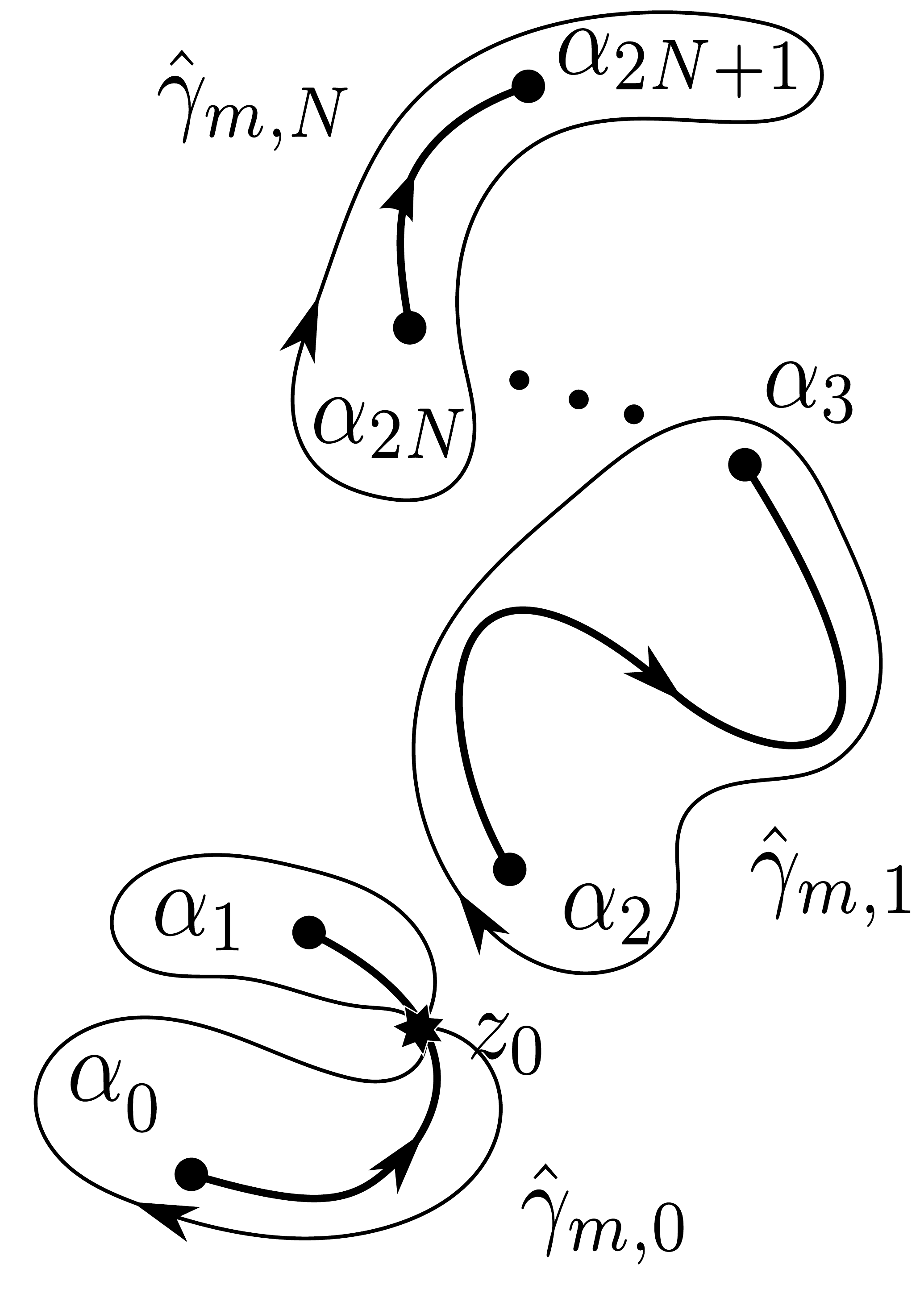}}
\subfigure[Contours $\ggh_{c,j}$ ]{\includegraphics[height=6cm]{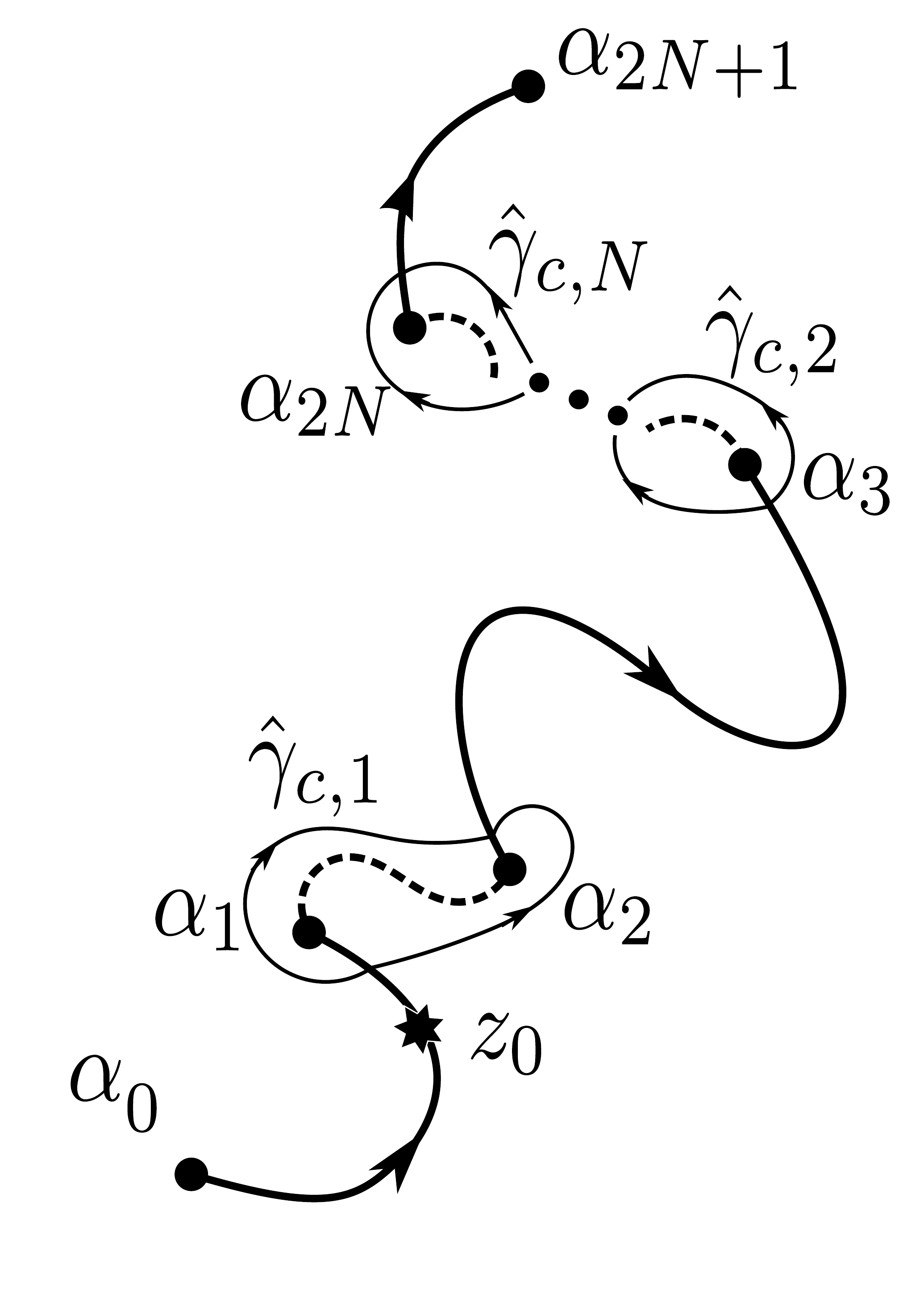}}
\caption{\label{fig:loop_contours} Contours of integration for functions $g(z)$ and $h(z)$ (\ref{eq:g(z)}) and (\ref{eq:h(z)}).
$z_0$ is a point of non-analyticity of $f(z)$ on $\gamma$, and $z_0=z_0(\beta_1)$ depends of an external parameter $\beta_1$.}
\end{center}
\end{figure}

A bounded function $g(z)$ satisfying these conditions
can be written explicitly, with the aid of the radical
\begin{equation}
R(z)=
\left({\prod_{j=0}^{2N+1}\left(z-\alpha_j\right)}\right)^{\frac{1}{2}},
\ \ \ \mbox{where}
\lim_{z\to\infty}\frac{R(z)}{z^{N+1}}=-1,
\end{equation}
in which the main arcs are the branchcuts. To write the formula for $g$ we introduce
some notation.
 The main arcs are labeled $\g_{m,j}=(\a_{2 j}, \a_{2j+1}) ,j = 0, 1, . . . , N$.  The complementary arcs, are labeled
$\g_{c,j}=(\a_{2 j-1},  \a_{2 j}), j = 1, 2, . . . , N$.
The subscript $+$  in $R(\zeta)_+$ indicates that  the value of $R$
on the branchcut (that is,  on the contour) is taken as the
limiting value from the left side of the contour.
The following $g(z)$ (see \cite{TVZzero}) is analytic and
bounded in $\Cbar \backslash\g$ and satisfies the jump conditions \eqref{jumpmain} and \eqref{jumpcomp}
\begin{equation}
g(z)=\frac{R(z)}{2\pi i}
\left[\oint_{\g}\frac{f(\zeta)}{(\zeta-z)R(\zeta)_+}d\zeta +
\sum_{j=0}^{N}\oint_{\g_{m,j}}\frac{W_j}{(\zeta-z)R(\zeta)_+}d\zeta+
\sum_{j=1}^{N}\oint_{\g_{c,j}}\frac{\Omega_j}{(\zeta-z)R(\zeta)_+}d\zeta \right].
\label{eq:g(z)}
\end{equation}
The real constants $W_j$ and $\Omega_j$ are chosen so that $g(z)$ is
$O(1)$ as $z\to\infty$, and, hence, analytic at $z=\infty$. They are
calculated as the unique solution to a linear system of
moment conditions that guarantee this  (see equations (14) of paper \cite{TV_det}).

Without loss of generality, we can fix one of the constants $W_j$, say $W_0=0$.
If $W_0\neq 0$, this can be archived by recalibrating the RH problem by considering
$g(z)-\frac{1}{2}W_0$ and subtracting $W_0$ to $g(\infty)$ and all $W_j$ on main arcs
$\gg_{m,j}$.

Rather than to perform integration along arcs we prefer to integrate along loops
surrounding these arcs. Thus, we define the function $h(z)$
\begin{equation}
h(z)=\frac{R(z)}{2\pi i}
\left[\oint_{\ggh}\frac{f(\zeta)}{(\zeta-z)R(\zeta)}d\zeta +
\sum_{j=0}^{N}\oint_{\ggh_{m,j}}\frac{W_j}{(\zeta-z)R(\zeta)}d\zeta+
\sum_{j=1}^{N}\oint_{\ggh_{c,j}}\frac{\Omega_j}{(\zeta-z)R(\zeta)}d\zeta \right],
\label{eq:h(z)}
\end{equation}
in which the oriented loop $\ggh$ surrounds contour $\gg$, $\ggh_{c,j}$ encircles
the complementary arc $\g_{c,j}$ (notice different orientations of the two
halves of the loop), and $\ggh_{m,j}$ encircles the main arc $\g_{m,j}$.
All loops are oriented clockwise, $z$ lies inside the contour $\ggh$ and
outside the loops $\ggh_{m,j}$ and $\ggh_{c,j}$.

Passing from arcs to loops introduces factor of $2$ and $z$ cutting through the
deforming loop around $\gg$ introduces a residue in the first integral.

Thus the relation of $g(z)$ and $h(z)$ is,
\begin{equation}
g(z)=\frac{1}{2}(h(z)+f(z)).
\end{equation}

The jump conditions for $h$ along the contour  are simpler than the
jumps of $g(z)$, thus, $h(z)$ is the most natural object in the
Riemann-Hilbert analysis. Results are easily reformulated in terms $g(z)$,
whose main advantage is its analyticity off the contour. The analyticity
of $h$ of the contour, is compromised at the non-analytic points of $f$
and on the real axis.

In order to evaluate the  limits of function $h(z)$
at a branchpoint, coming from the left or from the right
side of the RHP contour, we allow $z$ to cross inside the loops
of the main and complementary arcs adjacent to the branchpoint and we
correct by introducing the corresponding residue.
Denoting  the expression in the bracket
for this positioning of $z$ by
\begin{equation}
B(z)=\oint_{\ggh}\frac{f(\zeta)}{(\zeta-z)R(\zeta)}d\zeta +
\sum_{j=0}^{N}\oint_{\ggh_{m,j}}\frac{W_j}{(\zeta-z)R(\zeta)}d\zeta+
\sum_{j=1}^{N}\oint_{\ggh_{c,j}}\frac{\Omega_j}{(\zeta-z)R(\zeta)}d\zeta,
\label{eq:B(z)}
\end{equation}
we obtain, for $B$ calculated with $z$ {\it inside} the two loops,
\begin{equation}
h(z)=W_j \pm\O_j+\frac{R(z)B(z)}{2\pi i} \ \ \mbox{or} \ \ h(z)=W_j\pm\O_{j+1}+\frac{R(z)B(z)}{2\pi i},
\end{equation}
where the first equation applies near the branchpoint $\a_{2j}$,
the second  equation applies near the branchpoint $\a_{2j+1}$,
and $+$ or $-$ applies when  $z$ is to the left or right of the
contour respectively. The $\pm$ term is zero at the endpoints
of the contour $\a_0$ and $\a_{2N+1}$.

It is important to notice
that $B(z)$ is analytic at $z=\a_j$ and can thus be expanded to a
 convergent power series in the neighborhood of $\a_j$,
\begin{equation}
 B(z)=\nu_{1,j}+\nu_{2,j}(z-\a_j)+\nu_{3,j}(z-\a_j)^2+\ldots,
\end{equation}
 with coefficients that are analytic with respect to all of the branchpoints.

The growth/decay properties of $e^{\pm ih(z)/\eps}$ near the RHP contour $\g$
is a crucial to the semiclassical
analysis of matrix RHP. The requirement that  the constants $W_i$ and
$\O_i$ be real is a fall-out of this. An additional fall-out is the following
{\it sign conditions} for $h$ \cite{TVZzero}
\begin{equation}
 \Im h=0 \ \ \mbox{on main arcs};
\ \ \Im h<0 \ \ \mbox{left and right of main arcs}
\end{equation}
\begin{equation}
 \Im h>0 \ \ \mbox{on complementary arcs}
\end{equation}
The last condition must also apply on two ``extension arcs'', or
arcs extending the contour to $+\infty$ on one side and to
$-\infty$ on the other.

For the sign conditions to be satisfied, it is necessary,
but not sufficient that
\begin{equation}
 B(\a_j)=0, \ \ \ j=0,1,2,\ldots, 2N+1,
 \label{eq:mod_B}
\end{equation}
equivalently,
\begin{equation}
 \nu_{1,j}(\a_j)=0, \ \ \ j=0,1,2,\ldots, 2N+1.
\end{equation}
equivalently,
\begin{equation}
 h(z)=\mbox{analytic}+ O((z-\a_j)^{\frac{3}{2}}), \ \mbox{as} \  z\to\a_j, \ \ j=0,1,2,\ldots, 2N+1.
\end{equation}
equivalently,
\begin{equation}
 g(z)=\mbox{analytic}+ O((z-\a_j)^{\frac{3}{2}}), \ \mbox{as} \  z\to\a_j, \ \ j=0,1,2,\ldots, 2N+1.
\end{equation}
equivalently,
\begin{equation}
 h'(\a_j)=0,  \ \ j=0,1,2,\ldots, 2N+1.
\end{equation}

\begin{figure}
\begin{center}
{\includegraphics[height=8cm]{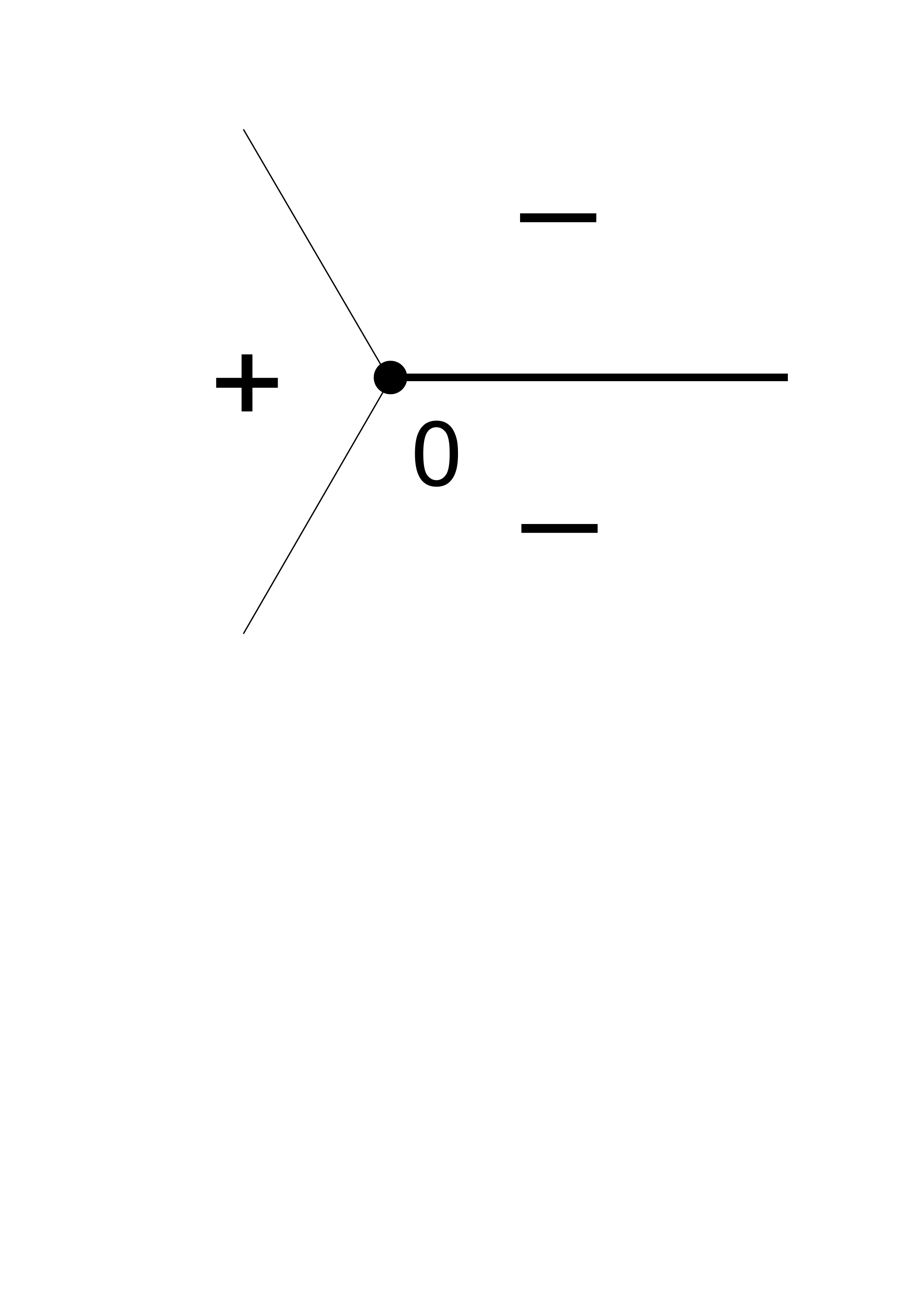}}
\caption{\label{fig:sqrt_signs} Sign structure of $\Im\sqrt z$ near the origin where
the branchcut is chosen along the positive semiaxis.}
\end{center}
\end{figure}
The necessity of the coefficient of the linear term of $B$ being
zero is intimately related to a simple fact about $\sqrt z$,  with
the positive semiaxis as a branchcut playing the role of a main arc.
With a sign determination that gives $\Im\sqrt z<0$ on
both sides of the branchcut, as required by the sign condition,
$\Im\sqrt z$ never turns positive, making the connection to
a complementary arc impossible. Figure \ref{fig:sqrt_signs} describes
how the sign conditions can be satisfied with $\sqrt{z}$.

In \cite{TV_det}, $B(z)$ is expressed as a ratio of two determinants through,
\begin{equation}
B(z)=\frac{2\pi i}{D}K(z),
\end{equation}
where

\begin{equation}
K(z)=\frac{1}{2\pi i}\left|\begin{array}{ccccccc}
\oint_{\ggh_{m,1}}\frac{d\zeta}{R(\zeta)} & \ldots & \oint_{\ggh_{m,1}}\frac{\zeta^{N-1}d\zeta}{R(\zeta)} &
\overline{\oint_{\ggh_{m,1}}\frac{d\zeta}{R(\zeta)}} & \ldots & \overline{\oint_{\ggh_{m,1}}\frac{\zeta^{N-1}d\zeta}{R(\zeta)}} & \oint_{\ggh_{m,1}}\frac{d\zeta}{(\zeta-z)R(\zeta)} \\
\ldots & \ldots & \ldots & \ldots & \ldots & \ldots & \ldots \\
\oint_{\ggh_{m,N}}\frac{d\zeta}{R(\zeta)} & \ldots & \oint_{\ggh_{m,N}}\frac{\zeta^{N-1}d\zeta}{R(\zeta)} &
\overline{\oint_{\ggh_{m,N}}\frac{d\zeta}{R(\zeta)}} & \ldots & \overline{\oint_{\ggh_{m,N}}\frac{\zeta^{N-1}d\zeta}{R(\zeta)}} &
\oint_{\ggh_{m,N}}\frac{d\zeta}{(\zeta-z)R(\zeta)} \\
\oint_{\ggh_{c,1}}\frac{d\zeta}{R(\zeta)} & \ldots & \oint_{\ggh_{c,1}}\frac{\zeta^{N-1}d\zeta}{R(\zeta)} &
\overline{\oint_{\ggh_{c,1}}\frac{d\zeta}{R(\zeta)}} & \ldots & \overline{\oint_{\ggh_{c,1}}\frac{\zeta^{N-1}d\zeta}{R(\zeta)}} &
\oint_{\ggh_{c,1}}\frac{d\zeta}{(\zeta-z)R(\zeta)} \\
\ldots & \ldots & \ldots & \ldots & \ldots & \ldots & \ldots \\
\oint_{\ggh_{c,N}}\frac{d\zeta}{R(\zeta)} & \ldots & \oint_{\ggh_{c,N}}\frac{\zeta^{N-1}d\zeta}{R(\zeta)} &
\overline{\oint_{\ggh_{c,N}}\frac{d\zeta}{R(\zeta)}} & \ldots & \overline{\oint_{\ggh_{c,N}}\frac{\zeta^{N-1}d\zeta}{R(\zeta)}} &
\oint_{\ggh_{c,N}}\frac{d\zeta}{(\zeta-z)R(\zeta)} \\
\oint_{\ggh}\frac{f(\zeta)d\zeta}{R(\zeta)} & \ldots & \oint_{\ggh}\frac{\zeta^{N-1}f(\zeta)d\zeta}{R(\zeta)} &
\overline{\oint_{\ggh}\frac{f(\zeta)d\zeta}{R(\zeta)}} & \ldots & \overline{\oint_{\ggh}\frac{\zeta^{N-1}f(\zeta)d\zeta}{R(\zeta)}} &
\oint_{\ggh}\frac{f(\zeta)d\zeta}{(\zeta-z)R(\zeta)}
\end{array}\right|
\label{eq:K_function}
\end{equation}
and
\begin{equation}
D=\left|\begin{array}{cccccc}
\oint_{\ggh_{m,1}}\frac{d\zeta}{R(\zeta)} & \ldots & \oint_{\ggh_{m,1}}\frac{\zeta^{N-1}d\zeta}{R(\zeta)} &
\overline{\oint_{\ggh_{m,1}}\frac{d\zeta}{R(\zeta)}} & \ldots & \overline{\oint_{\ggh_{m,1}}\frac{\zeta^{N-1}d\zeta}{R(\zeta)}} \\
\ldots & \ldots & \ldots & \ldots & \ldots & \ldots \\
\oint_{\ggh_{m,N}}\frac{d\zeta}{R(\zeta)} & \ldots & \oint_{\ggh_{m,N}}\frac{\zeta^{N-1}d\zeta}{R(\zeta)} &
\overline{\oint_{\ggh_{m,N}}\frac{d\zeta}{R(\zeta)}} & \ldots & \overline{\oint_{\ggh_{m,N}}\frac{\zeta^{N-1}d\zeta}{R(\zeta)}}  \\
\oint_{\ggh_{c,1}}\frac{d\zeta}{R(\zeta)} & \ldots & \oint_{\ggh_{c,1}}\frac{\zeta^{N-1}d\zeta}{R(\zeta)} &
\overline{\oint_{\ggh_{c,1}}\frac{d\zeta}{R(\zeta)}} & \ldots & \overline{\oint_{\ggh_{c,1}}\frac{\zeta^{N-1}d\zeta}{R(\zeta)}} \\
\ldots & \ldots & \ldots & \ldots & \ldots & \ldots  \\
\oint_{\ggh_{c,N}}\frac{d\zeta}{R(\zeta)} & \ldots & \oint_{\ggh_{c,N}}\frac{\zeta^{N-1}d\zeta}{R(\zeta)} &
\overline{\oint_{\ggh_{c,N}}\frac{d\zeta}{R(\zeta)}} & \ldots & \overline{\oint_{\ggh_{c,N}}\frac{\zeta^{N-1}d\zeta}{R(\zeta)}}
\end{array}\right|.
\label{D_det}
\end{equation}
and the following relations are shown
\begin{equation}
g(z)=\frac{R(z)}{2D}K(z),
\end{equation}
where $z$ lies outside of $\ggh$ and outside all $\ggh_{c,j}$ and $\ggh_{m,j}$ and
\begin{equation}
h(z)=\frac{R(z)}{D}K(z),
\label{eq:h_and_K}
\end{equation}
where $z$ lies inside of $\ggh$ and outside all $\ggh_{c,j}$ and $\ggh_{m,j}$.

For the rest of the paper we consider function $K(z)$ which is a constant multiple of $B(z)$.
From (\ref{eq:mod_B}) the modulation equations to determine the branchpoints $\alpha_j$ are the following
\begin{equation}
K(\alpha_j)=0,\ \ \ j=0,\ldots,2N+1.
\label{eq:mod_K}
\end{equation}

\section{Smooth parametric dependence of the branchpoints $\alpha_j$}

By external parameters $\vb=(\beta_1,\beta_2,\ldots)$ we think either $\vb\in\mathbb{R}^m$ or $\vb\in\mathbb{C}^m$,
 $m\ge 1$.

\begin{definition}$ $\\
Let a simple contour $\gamma_0$ consist of a finite union of finite length oriented simple arcs
in the complex plane $\gamma_0=\left(\cup\gg_{m,j}\right)\cup \left(\cup\gg_{c,j}\right)$ with the
distinct end points $\va_0=\left\{\alpha_j\right\}^{2N+1}_{j=0}$, as shown in Fig. \ref{fig:RHP_contours}.
For a fixed vector of external parameters $\vb_0$, we define
\[
\mathbb{L}(\gamma_0,\va_0,\vb_0)
\]
to be the set of functions $f(z,\vb)$ defined in some open neighborhood of $(\gg_0,\vb_0)$ which have the form
\begin{equation}
f(z,\vb)=c(\vb)\left(z-z_0\right)\log\left(z-z_0\right)+A(z,\vb),
\label{eq:f_form_def}
\end{equation}
with $z_0\in \gamma_{m,j} \setminus \cup_k \left\{ \alpha_k\right\}$ for some $j$
(without loss of generality, let $z_0\in \gg_{m,0}\setminus \left\{\alpha_0,\alpha_1\right\}$);
and where any branch of the logarithm is chosen so that the branchcut does not intersect $\gamma_0$.
Assume $c(\vb)$ and $A(z,\vb)$ are twice continuously differentiable in parameters $\vb$ in some open neighborhood of $\vb_0$;
and for each $\vb$ in some open neighborhood of $\vb_0$, $A(z,\vb)$ is analytic in $z$ in some
open neighborhood of $\gg_0$.

Let $z_0$, the point of logarithmic singularity of $f$, depend on one of the parameters,
say $\beta_1$. Assume $z_0=z_0(\beta_1)$ is continuously differentiable in $\beta_1$.
For simplicity, assume that $z_0(\beta_1+\Delta\beta_1)$ approaches to $\gamma_0$
as $\Delta\beta_1\to 0$ non-tangentially.

\label{def:3.1}
\end{definition}

Note that for fixed $\va_0$ and $\vb_0$ the contour $\gamma_0$ aside from passing
through $\va_0$ and $z_0$ is free to deform continuously as long as
$f\in\mathbb{L}(\gamma_0,\va_0,\vb_0)$. This condition fixes the (distinct)
end points of $\gamma_0$ and the singularity $z_0$. Thus a contour $\gamma_0$
from the definition \ref{def:3.1} depends on the end points $\va_0$ and on
$\beta^0_1$ through $z_0$, $\gamma(\va_0,\beta^0_1)=\gamma_0$.

\begin{lemma}
Let $\gamma_0$ be a simple oriented contour with the distinct end points $\va_0$
and assume $f\in\mathbb{L}(\gamma_0,\va_0,\vb_0)$.
Then there exist open neighborhoods of $\va_0$ and $\vb_0$
such that
\[
f\in\mathbb{L}(\gamma,\va,\vb)
\]
for all $\va$ in the neighborhood of $\va_0$, for all $\vb=(\beta_1,\ldots)$ in the neighborhood of $\vb_0$,
and some contour $\gamma=\gamma(\va, \beta_1)$.
\label{lem:f_in_L_near}
\end{lemma}

\begin{proof}
From the definition $f\in\mathbb{L}(\gamma_0,\va_0,\vb_0)$ implies that $f\in\mathbb{L}(\gamma_0,\va_0,\vb)$
for all $\vb$ in some open neighborhood of $\vb_0$.

Now fix some $\vb$ in the neighborhood of $\vb_0$ with $f\in\mathbb{L}(\gamma_0,\va_0,\vb)$.
Since the form of $f$ (\ref{eq:f_form_def}) is valid in some neighborhood of $\gamma_0$,
then there is a neighborhood of $\va_0$ where for all $\va$ the contour
$\gamma_0$ with the end points $\va_0$ can be deformed into $\gamma$ with the end points $\va$.
For example, by continuously connecting the end points $\va_0$ with the points $\va$.
Thus $f\in\mathbb{L}(\gamma,\va,\vb)$.

The size of the neighborhoods of $\vb_0$ and $\va_0$ is determined by the distance
$\alpha_j$'s with respect to each other and the distance between $\gamma$ and the
singularities of $f(z)$ (other than $z_0$).

\end{proof}

By considering the loop contours $\ggh$, $\ggh_{m,j}$, $\ggh_{c,j}$ (see Figure \ref{fig:loop_contours})
the explicit dependence of the contours of integration on the end points $\va$ is removed
(for example in (\ref{eq:Lemma3.2_int1}-\ref{eq:Lemma3.2_int5-6})). So even though
$\gamma=\gamma(\va,\beta_1)$, in all our evaluations below $\ggh=\ggh(\beta_1)$.

The main difficulty is the dependence of $f(z)$ (thus the RHP (\ref{jumpmain})) and the modulation
equations (\ref{eq:mod_K}) on parameter $\beta_1$ which also controls the point $z_0$ on the
contour $\ggh$. We show that the dependence on $\beta_1$ (moreover on $\vb$) is smooth.

\begin{remark}
In \cite{TV_det} was considered the case when the contour $\gamma$ was
independent of external parameters $\vb$ which led to simpler conditions on $f$ being continuous
on $\gamma$ except at finitely many points. If $f\in \mathbb{L}(\gamma,\va,\vb)$, then all results
below for $\beta_2,\beta_3,\ldots$ follow from \cite{TV_det}. However the harder
case of the dependence on $\beta_1$ is new.
\end{remark}

Consider the system of modulation equations as a function of both $\va$ and $\vb$
\begin{equation}
K(\alpha_j)=K(\alpha_j,\va,\vb)=0, \ \ \ j=0,1,..., 2N+1. \label{K_eqn}
\end{equation}

Note, that the function $K(z,\va,\vb)$ is defined by (\ref{eq:K_function}) and through (\ref{eq:h_and_K}) defines $h(z,\va,\vb)$,
which satisfies a scalar Riemann-Hilbert problem on $\gamma=\gamma(\va,\beta_1)$:
\begin{equation}
\left\{
\begin{array}{l}
h_{+}(z)+h_{-}(z)=2W_j(\va,\vb),\ \mbox{on} \ \gg_{m,j}, \ \ j=0,1,...,N,\\
h_{+}(z)-h_{-}(z)=2\Omega_j(\va,\vb),\ \mbox{on} \ \gg_{c,j}, \ \ j=1,...,N,\\
h(z)+f(z)\ \mbox{is analytic in}\ \mathbb{\overline{C}}\backslash\gamma.
\end{array}
\right.
\label{eq:first_time_RHP_h_at_vb}
\end{equation}
So every time we use $K$ we understand that there is an underlying scalar Riemann-Hilbert problem
which depends on $\va$ and $\vb$.

Following \cite{TV_det}, we differentiate (\ref{K_eqn}) with respect to $\beta_k$
\begin{equation}
\sum_{l}\frac{\partial K(\alpha_j)}{\partial\alpha_l}\frac{\partial\alpha_l}{\partial\beta_k}+\frac{\partial K(\alpha_j)}{\partial\beta_k}=0,
\end{equation}
where the matrix $\left\{\frac{\partial K(\alpha_j)}{\partial\alpha_l}\right\}_{j,l}$ is diagonal \cite{TV_det} so
\begin{equation}
\frac{\partial K(\alpha_j)}{\partial\alpha_j}\frac{\partial\alpha_j}{\partial\beta_k}=-\frac{\partial K(\alpha_j)}{\partial\beta_k}.
\end{equation}
Since
\begin{equation}
\frac{\partial K(\alpha_j)}{\partial\alpha_j}=\frac{D(\va,\vb)}{2\pi i}\oint_{\ggh(\beta_1)}\frac{f'(\zeta,\vb)}{(\zeta-\alpha_j)R(\zeta,\va)}d\zeta
\end{equation}
we arrive to the evolution equations for $\alpha_j$:
\begin{equation}
\frac{\partial\alpha_j}{\partial\beta_k}=-\frac{2\pi i \frac{\partial K(\alpha_j)}{\partial\beta_k}}{D(\va,\vb) \oint_{\ggh(\beta_1)}\frac{f'(\zeta,\vb)}{(\zeta-\alpha_j)R(\zeta,\va)}d\zeta}.
\label{eq_alpha_mu}
\end{equation}

Since $D\neq 0$ for distinct $\alpha_j$'s \cite{TV_det}, next we need to estimate the partial derivatives $\frac{\partial K(\alpha_j)}{\partial\beta_k}$.

\begin{equation}
\frac{\partial K(\alpha_j)}{\partial\beta_k}=
\frac{1}{2\pi i}\frac{\partial }{\partial\beta_k}
\left|\begin{array}{ccccccc}
\oint_{\ggh_{m,1}}\frac{d\zeta}{R(\zeta)} & \ldots & \oint_{\ggh_{m,1}}\frac{\zeta^{N-1}d\zeta}{R(\zeta)} &
\overline{\oint_{\ggh_{m,1}}\frac{d\zeta}{R(\zeta)}} & \ldots & \overline{\oint_{\ggh_{m,1}}\frac{\zeta^{N-1}d\zeta}{R(\zeta)}} & \oint_{\ggh_{m,1}}\frac{d\zeta}{(\zeta-\alpha_j)R(\zeta)} \\
\ldots & \ldots & \ldots & \ldots & \ldots & \ldots & \ldots \\
\oint_{\ggh_{m,N}}\frac{d\zeta}{R(\zeta)} & \ldots & \oint_{\ggh_{m,N}}\frac{\zeta^{N-1}d\zeta}{R(\zeta)} &
\overline{\oint_{\ggh_{m,N}}\frac{d\zeta}{R(\zeta)}} & \ldots & \overline{\oint_{\ggh_{m,N}}\frac{\zeta^{N-1}d\zeta}{R(\zeta)}} &
\oint_{\ggh_{m,N}}\frac{d\zeta}{(\zeta-\alpha_j)R(\zeta)} \\
\oint_{\ggh_{c,1}}\frac{d\zeta}{R(\zeta)} & \ldots & \oint_{\ggh_{c,1}}\frac{\zeta^{N-1}d\zeta}{R(\zeta)} &
\overline{\oint_{\ggh_{c,1}}\frac{d\zeta}{R(\zeta)}} & \ldots & \overline{\oint_{\ggh_{c,1}}\frac{\zeta^{N-1}d\zeta}{R(\zeta)}} &
\oint_{\ggh_{c,1}}\frac{d\zeta}{(\zeta-\alpha_j)R(\zeta)} \\
\ldots & \ldots & \ldots & \ldots & \ldots & \ldots & \ldots \\
\oint_{\ggh_{c,N}}\frac{d\zeta}{R(\zeta)} & \ldots & \oint_{\ggh_{c,N}}\frac{\zeta^{N-1}d\zeta}{R(\zeta)} &
\overline{\oint_{\ggh_{c,N}}\frac{d\zeta}{R(\zeta)}} & \ldots & \overline{\oint_{\ggh_{c,N}}\frac{\zeta^{N-1}d\zeta}{R(\zeta)}} &
\oint_{\ggh_{c,N}}\frac{d\zeta}{(\zeta-\alpha_j)R(\zeta)} \\
\oint_{\ggh}\frac{f(\zeta)d\zeta}{R(\zeta)} & \ldots & \oint_{\ggh}\frac{\zeta^{N-1}f(\zeta)d\zeta}{R(\zeta)} &
\overline{\oint_{\ggh}\frac{f(\zeta)d\zeta}{R(\zeta)}} & \ldots & \overline{\oint_{\ggh}\frac{\zeta^{N-1}f(\zeta)d\zeta}{R(\zeta)}} &
\oint_{\ggh}\frac{f(\zeta)d\zeta}{(\zeta-\alpha_j)R(\zeta)}
\end{array}\right|.
\label{eq:K_function_pre_lemma}
\end{equation}

To proceed we need the following technical lemma.

\begin{lemma}$ $\\

Let $f\in\mathbb{L}(\gamma,\va,\vb_0)$, where the contour $\gamma=\gamma(\va,\beta_1)$ has fixed end points $\va$.
Then there is an open neighborhood of $\vb_0$ such that for all $\vb$ in the neighborhood of $\vb_0$

\begin{equation}
\frac{\partial }{\partial\beta_k}\oint_{\ggh(\beta_1)}\frac{\zeta^n f(\zeta,\bb)d\zeta}{R(\zeta,\va)}
=\oint_{\ggh(\beta_1)}\frac{\zeta^n \frac{\partial f(\zeta,\bb)}{\partial\beta_k}d\zeta}{R(\zeta,\va)},\ \ n\in\N,
\label{eq:Lemma3.2_int1}
\end{equation}

\begin{equation}
\frac{\partial }{\partial\beta_k} \oint_{\ggh(\beta_1)}\frac{f(\zeta,\bb)d\zeta}{(\xi-\alpha_j)R(\zeta,\va)}
=\oint_{\ggh(\beta_1)}\frac{\frac{\partial f(\zeta,\bb)}{\partial\beta_k}d\zeta}{(\xi-\alpha_j)R(\zeta,\va)}, \ \ n\in\N,
\label{eq:Lemma3.2_int2}
\end{equation}

\begin{equation}
\frac{\partial }{\partial\beta_k} \oint_{\ggh_{m,j}}\frac{\zeta^n d\zeta}{R(\zeta,\va)}=0, \ \ \
\frac{\partial }{\partial\beta_k} \oint_{\ggh_{m,j}}\frac{d\zeta}{(\xi-\alpha_j)R(\zeta,\va)}=0, \ \ j=1,2,\ldots,N,\ n\in\N,
\label{eq:Lemma3.2_int3-4}
\end{equation}

\begin{equation}
\frac{\partial }{\partial\beta_k} \oint_{\ggh_{c,j}}\frac{\zeta^n d\zeta}{R(\zeta,\va)}=0, \ \ \
\frac{\partial }{\partial\beta_k} \oint_{\ggh_{c,j}}\frac{d\zeta}{(\xi-\alpha_j)R(\zeta,\va)}=0, \ \ j=1,2,\ldots,N,\ n\in\N,
\label{eq:Lemma3.2_int5-6}
\end{equation}

for $k\ge 1$.

\label{lem:par_int_zero}

\end{lemma}

\begin{proof}

The idea of the proof is to consider the finite differences and take the limit. The main difficulty is the case
$k=1$ when both the integrand and the contour depend on $\beta_1$.

Denote the integral on the left in (\ref{eq:Lemma3.2_int1}) as $I_1$
\begin{equation}
I_1(\vb)=\oint_{\ggh(\beta_1)}\frac{\zeta^n f(\zeta,\bb)d\zeta}{R(\zeta,\va)}.
\end{equation}

First, consider $k\ge 2$. Without loss of generality, assume that the branchcut of logarithm in $f(z,\vb)$ is chosen from
$z_0\left(\beta_1\right)$ horizontally (along the straight line $z=i\Im z_0$) to the right
near $z_0$ (see Fig. \ref{fig:log_branchcuts_k}).

The contour of integration $\ggh$ near $z_0$ is pushed to the logarithmic branchcut of $f$.
Fix complex points $\delta_1$ and $\delta_2$ near $z_0$ with $\Im\delta_1=\Im\delta_2=\Im z_0$ (see Fig. \ref{fig:log_branchcuts_k}).
Then $\ggh$ is split into
two parts: $[\delta_1,\delta_2]$ and its complement. Across the logarithmic branchcut $\left[z_0,\delta_2\right]$, $f(z,\bb)$
has a jump $2\pi i \left(z_0-z\right)c(\vec\beta)$ and no jump across $\left[\delta_1,z_0\right]$.

\begin{figure}
\begin{center}
\includegraphics[height=5cm]{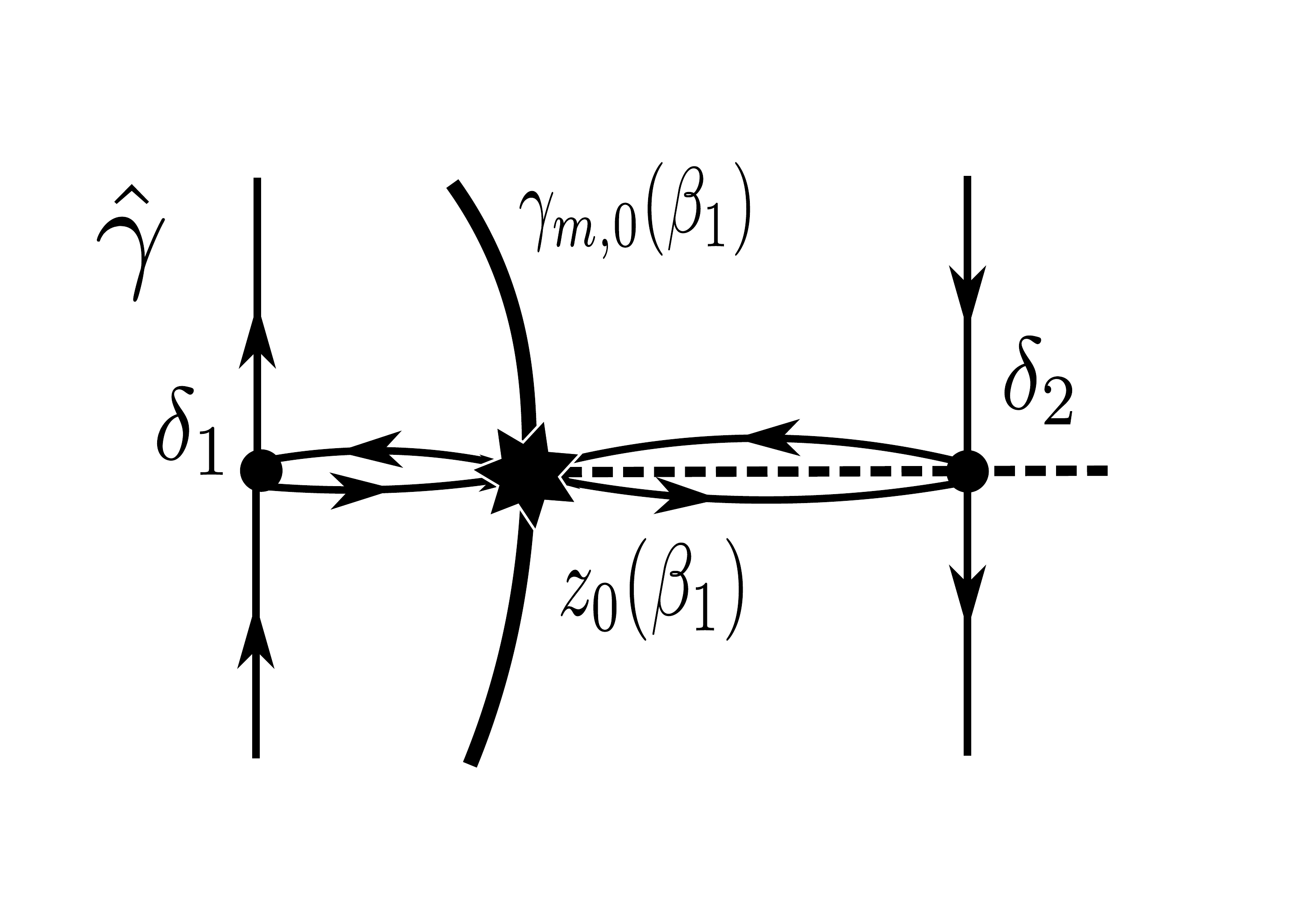}
\includegraphics[height=5cm]{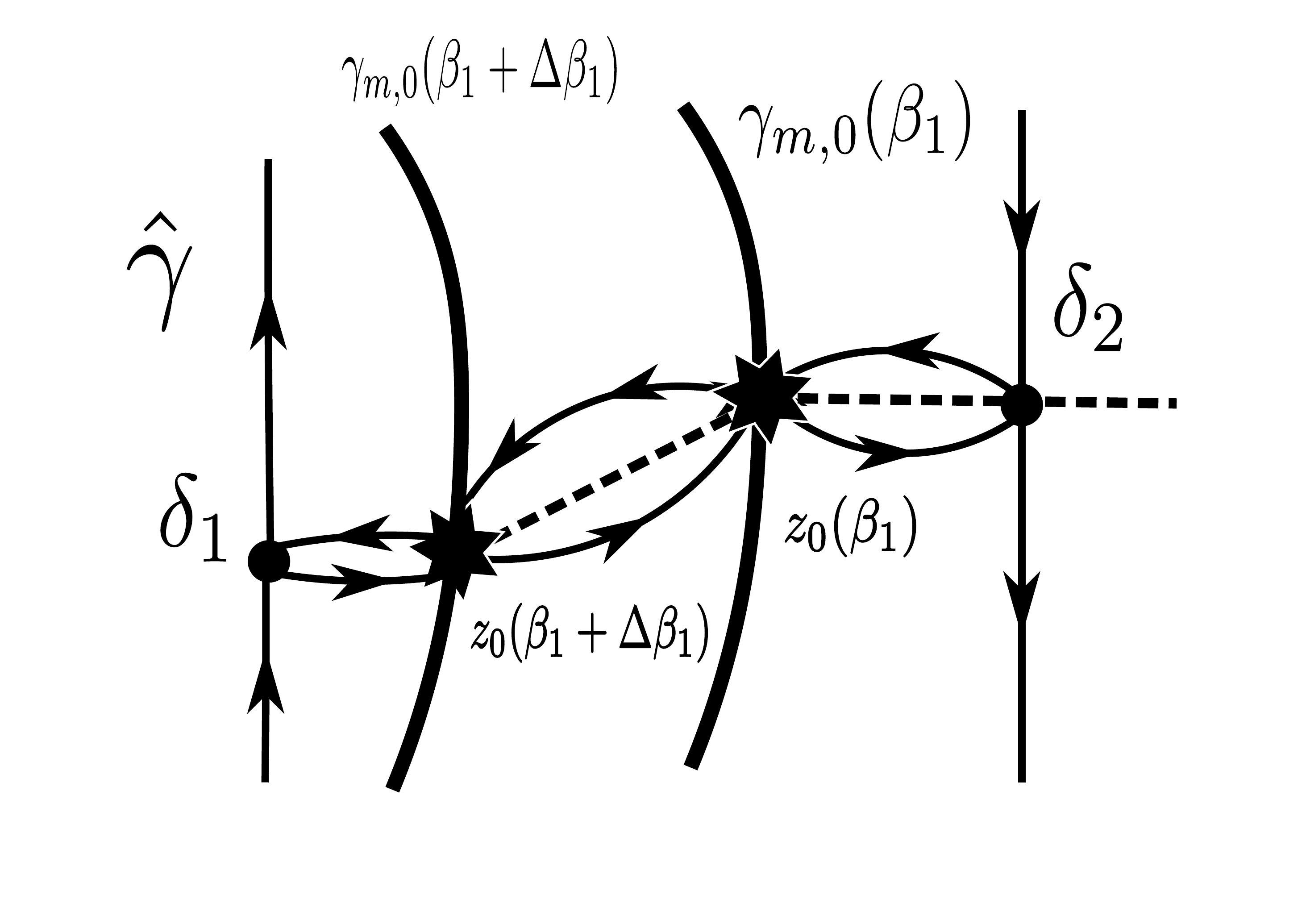}
\caption{\label{fig:log_branchcuts_k}Deforming the contours of integration for $I_1$ near $z_0$.
The dashed line is the logarithmic branchcut of $\log(z-z_0)$ in $f$ and
$\frac{\partial f(\zeta,\bb)}{\partial\beta_k}$ in the integrals
(\ref{eq:Lemma3.2_int1}-\ref{eq:Lemma3.2_int2}).}
\end{center}
\end{figure}

Take a small $\Delta\beta_k$ and consider finite differences
\begin{equation}
\frac{I_1(\vb+\Delta \vb_k)-I_1(\bb)}{\Delta \beta_k}
=\oint_{\ggh(\beta_1)}\frac{\zeta^n \frac{f(\zeta,\vb+\Delta \vb_k) -f(\zeta,\vb)}{\Delta \beta_k}}{R(\zeta,\va)}d\zeta,
\label{eq:I_1_beta_k}
\end{equation}
where $\Delta \vec\beta_k=\Delta \beta_k \vec{e}_k$ with $\vec{e}_k$ to be the standard basis
$\left(\vec{e}_k\right)_j=\left\{\begin{array}{l}1,\ j=k,\\0,\ j\neq k\end{array}\right.$.

The limit $\Delta\beta_k\to 0$ can be interchanged with the integral in (\ref{eq:I_1_beta_k})
since the integrand is uniformly bounded in $\beta_k$ on $\ggh$, $k\ge2$.

In the case $k=1$, there are two logarithmic branchcuts of $f(z,\vb)$ and $f(z,\vb+\vDb_1)$.
By assumption $z_0$ is not tangential to $\gamma$ and so for small $\Delta\beta_1$,
$z_0(\beta_1+\Delta\beta_1)$ is not on $\gamma$. Assume that $z_0(\beta_1+\Delta\beta_1)$ is
on the left (positive) side of $\gamma_{m,0}(\beta_1)$. The case when $z_0(\beta_1+\Delta\beta_1)$ is
on the other side of $\gamma_{m,0}(\beta_1)$ is done similarly.

Without loss of generality,
assume that the branchcut of logarithm in $f(z,\vb)$ is chosen as before and in $f(z,\vb+\vDb_1)$
the branchcut is chosen from $z_0(\beta_1+\Delta\beta_1)$
to $z_0\left(\beta_1\right)$ and horizontally to the right (see Fig. \ref{fig:log_branchcuts_k}). Then similarly,
we choose fixed points $\delta_1$ and $\delta_2$ so that $\Im\delta_1=\Im z_0\left(\beta_1+\Delta\beta_1\right)$
and $\Im\delta_2=\Im z_0\left(\beta_1\right)$.
The contour of integration $\ggh$ is pushed to the logarithmic branchcut near $z_0$ and split into $\left[\delta_1,\delta_2\right]$
and its complement, where by $\left[\delta_1,\delta_2\right]$ we understand
$\left[\delta_1,z_0\left(\beta_1+\Delta\beta_1\right)\right]\cup[z_0\left(\beta_1+\Delta\beta_1\right),z_0\left(\beta_1\right)]\cup\left[z_0\left(\beta_1\right)\delta_2\right]$.

Across $\left[z_0(\beta_1),\delta_2\right]$, $f(z,\bb)$
has a jump $2\pi i \left(z_0(\beta_1)-z\right)c(\vec\beta)$.
Similarly, $f(z,\bb+\Delta\vec\beta_1)$ has a jump $2\pi i \left(z_0(\beta_1+\Delta\beta_1)-z\right)c(\vb+\vDb_1)$ on $[z_0\left(\beta_1+\Delta\beta_1\right),z_0\left(\beta_1\right)]\cup\left[z_0\left(\beta_1\right),\delta_2\right]$.

For $k=1$,
\begin{equation}
\frac{I_1(\vb+\Delta \vb_1)-I_1(\bb)}{\Delta \beta_1}
=\oint_{\ggh(\beta_1)\backslash[\delta_1,\delta_2]}\frac{\zeta^n \frac{f(\zeta,\vb+\Delta \vb_1)-f(\zeta,\bb)}{\Delta \beta_1}}{R(\zeta,\va)}d\zeta
\end{equation}

\begin{equation}
+\int_{[z_0(\beta_1),\delta_2]}\frac{\zeta^n \frac{2\pi i (z_0(\beta_1+\Delta\beta_1)-\zeta)c(\vb+\Delta\vb_1)-2\pi i (z_0(\beta_1)-\zeta)c(\vb)}{\Delta \beta_1}}{R(\zeta,\va)}d\zeta
\end{equation}

\begin{equation}
+\int_{[z_0(\beta_1+\Delta\beta_1),z_0(\beta_1)]}\frac{\zeta^n \frac{2\pi i (z_0(\beta_1+\Delta\beta_1)-\zeta)c(\vb+\Delta\vb_1)}{\Delta \beta_1}}{R(\zeta,\va)}d\zeta.
\end{equation}
The last integral is $O\left(\Delta\beta_1\right)$ which is observed by the change of variables $y=\zeta-z_0(\beta_1+\Delta\beta_1)$
\begin{equation}
\int_{[z_0(\beta_1+\Delta\beta_1),z_0(\beta_1)]}\frac{\zeta^n \frac{2\pi i (z_0(\beta_1+\Delta\beta_1)-\zeta)c(\vb+\Delta\vb_1)}{\Delta \beta_1}}{R(\zeta,\va)}d\zeta
\label{eq:estimating_O(beta_1)_integral}
\end{equation}
\begin{equation}
=-\frac{ 2\pi i c(\vb+\Delta\vb_1)}{\Delta\beta_1} \int_{[0,z_0(\beta_1)-z_0(\beta_1+\Delta\beta_1)]} \frac{(y+z_0(\beta_1+\Delta\beta_1))^ny}{R(y+z_0(\beta_1+\Delta\beta_1),\va)}d y
\end{equation}
\begin{equation}
=O\left(\frac{1}{\Delta\beta_1} \int_{[0,O\left(\Delta\beta_1\right)]}ydy \right)=O\left(\Delta\beta_1\right), \ \mbox{as}\ \Delta\beta_1\to 0.
\end{equation}
Thus
\begin{equation}
\frac{I_1(\bb+\Delta \vb_1)-I_1(\bb)}{\Delta \beta_1}
=\oint_{\ggh(\beta_1)}\frac{\zeta^n \frac{f(\zeta,\vb+\Delta \vb_1)-f(\zeta,\bb)}{\Delta \beta_1}}{R(\zeta,\va)}d\zeta +O\left(\Delta\beta_1\right).
\label{eq:I_1_with_O(beta_1)}
\end{equation}

The last step is to take the limit as $\Delta\beta_1\to 0$ and to interchange it with the integral. The contour
of integration is split into two:
a small neighborhood near $z_0$ and its complement. For the integral near $z_0$, we demonstrate
through an explicit calculation of a simplified integral in the appendix that the limit can be passed
under the integral. The integral over the second part of the contour has the integrand uniformly bounded
in $\beta_1$ since $\log(\xi-z_0(\beta_1))$ in $\frac{\partial f}{\partial\beta_1}$is uniformly bounded away from $z_0$, so the limit and the integral can be interchanged.
This completes the proof for the first integral (\ref{eq:Lemma3.2_int1}).

The second integral (\ref{eq:Lemma3.2_int2}) is done similarly. The rest of the integrals (\ref{eq:Lemma3.2_int3-4})-(\ref{eq:Lemma3.2_int5-6}) do not
depend on $\beta_k$ since the only dependence on $\vb$ sits in $z_0(\beta_1)\in\gamma_{m,0}$.

Thus for fixed $\va$ the integrals (\ref{eq:Lemma3.2_int3-4})-(\ref{eq:Lemma3.2_int5-6}) are independent of $\beta_k$.

\end{proof}

Thus from (\ref{eq:K_function_pre_lemma}) using Lemma \ref{lem:par_int_zero} for $k\ge 1$ we get
\begin{equation}
\frac{\partial K(\alpha_j)}{\partial\beta_k}(\va,\vb)=
\frac{1}{2\pi i}\left|\begin{array}{ccccccc}
\oint_{\ggh_{m,1}}\frac{d\zeta}{R(\zeta)} & \ldots & \oint_{\ggh_{m,1}}\frac{\zeta^{N-1}d\zeta}{R(\zeta)} &
\overline{\oint_{\ggh_{m,1}}\frac{d\zeta}{R(\zeta)}} & \ldots & \overline{\oint_{\ggh_{m,1}}\frac{\zeta^{N-1}d\zeta}{R(\zeta)}} & \oint_{\ggh_{m,1}}\frac{d\zeta}{(\zeta-\alpha_j)R(\zeta)} \\
\ldots & \ldots & \ldots & \ldots & \ldots & \ldots & \ldots \\
\oint_{\ggh_{m,N}}\frac{d\zeta}{R(\zeta)} & \ldots & \oint_{\ggh_{m,N}}\frac{\zeta^{N-1}d\zeta}{R(\zeta)} &
\overline{\oint_{\ggh_{m,N}}\frac{d\zeta}{R(\zeta)}} & \ldots & \overline{\oint_{\ggh_{m,N}}\frac{\zeta^{N-1}d\zeta}{R(\zeta)}} &
\oint_{\ggh_{m,N}}\frac{d\zeta}{(\zeta-\alpha_j)R(\zeta)} \\
\oint_{\ggh_{c,1}}\frac{d\zeta}{R(\zeta)} & \ldots & \oint_{\ggh_{c,1}}\frac{\zeta^{N-1}d\zeta}{R(\zeta)} &
\overline{\oint_{\ggh_{c,1}}\frac{d\zeta}{R(\zeta)}} & \ldots & \overline{\oint_{\ggh_{c,1}}\frac{\zeta^{N-1}d\zeta}{R(\zeta)}} &
\oint_{\ggh_{c,1}}\frac{d\zeta}{(\zeta-\alpha_j)R(\zeta)} \\
\ldots & \ldots & \ldots & \ldots & \ldots & \ldots & \ldots \\
\oint_{\ggh_{c,N}}\frac{d\zeta}{R(\zeta)} & \ldots & \oint_{\ggh_{c,N}}\frac{\zeta^{N-1}d\zeta}{R(\zeta)} &
\overline{\oint_{\ggh_{c,N}}\frac{d\zeta}{R(\zeta)}} & \ldots & \overline{\oint_{\ggh_{c,N}}\frac{\zeta^{N-1}d\zeta}{R(\zeta)}} &
\oint_{\ggh_{c,N}}\frac{d\zeta}{(\zeta-\alpha_j)R(\zeta)} \\
\oint_{\ggh}\frac{f_{\beta_k}(\zeta)d\zeta}{R(\zeta)} & \ldots & \oint_{\ggh}\frac{\zeta^{N-1}f_{\beta_k}(\zeta)d\zeta}{R(\zeta)} &
\overline{\oint_{\ggh}\frac{f_{\beta_k}(\zeta)d\zeta}{R(\zeta)}} & \ldots & \overline{\oint_{\ggh}\frac{\zeta^{N-1}f_{\beta_k}(\zeta)d\zeta}{R(\zeta)}} &
\oint_{\ggh}\frac{f_{\beta_k}(\zeta)d\zeta}{(\zeta-\alpha_j)R(\zeta)}
\end{array}\right|,
\label{eq:DK/D_beta}
\end{equation}

and for $k+m\ge 3$
\begin{equation}
\frac{\partial^2 K(\alpha_j)}{\partial\beta_k\partial\beta_m}(\va,\vb)=
\frac{1}{2\pi i}\left|\begin{array}{ccccccc}
\oint_{\ggh_{m,1}}\frac{d\zeta}{R(\zeta)} & \ldots & \oint_{\ggh_{m,1}}\frac{\zeta^{N-1}d\zeta}{R(\zeta)} &
\overline{\oint_{\ggh_{m,1}}\frac{d\zeta}{R(\zeta)}} & \ldots & \overline{\oint_{\ggh_{m,1}}\frac{\zeta^{N-1}d\zeta}{R(\zeta)}} & \oint_{\ggh_{m,1}}\frac{d\zeta}{(\zeta-\alpha_j)R(\zeta)} \\
\ldots & \ldots & \ldots & \ldots & \ldots & \ldots & \ldots \\
\oint_{\ggh_{m,N}}\frac{d\zeta}{R(\zeta)} & \ldots & \oint_{\ggh_{m,N}}\frac{\zeta^{N-1}d\zeta}{R(\zeta)} &
\overline{\oint_{\ggh_{m,N}}\frac{d\zeta}{R(\zeta)}} & \ldots & \overline{\oint_{\ggh_{m,N}}\frac{\zeta^{N-1}d\zeta}{R(\zeta)}} &
\oint_{\ggh_{m,N}}\frac{d\zeta}{(\zeta-\alpha_j)R(\zeta)} \\
\oint_{\ggh_{c,1}}\frac{d\zeta}{R(\zeta)} & \ldots & \oint_{\ggh_{c,1}}\frac{\zeta^{N-1}d\zeta}{R(\zeta)} &
\overline{\oint_{\ggh_{c,1}}\frac{d\zeta}{R(\zeta)}} & \ldots & \overline{\oint_{\ggh_{c,1}}\frac{\zeta^{N-1}d\zeta}{R(\zeta)}} &
\oint_{\ggh_{c,1}}\frac{d\zeta}{(\zeta-\alpha_j)R(\zeta)} \\
\ldots & \ldots & \ldots & \ldots & \ldots & \ldots & \ldots \\
\oint_{\ggh_{c,N}}\frac{d\zeta}{R(\zeta)} & \ldots & \oint_{\ggh_{c,N}}\frac{\zeta^{N-1}d\zeta}{R(\zeta)} &
\overline{\oint_{\ggh_{c,N}}\frac{d\zeta}{R(\zeta)}} & \ldots & \overline{\oint_{\ggh_{c,N}}\frac{\zeta^{N-1}d\zeta}{R(\zeta)}} &
\oint_{\ggh_{c,N}}\frac{d\zeta}{(\zeta-\alpha_j)R(\zeta)} \\
\oint_{\ggh}\frac{f_{\beta_k\beta_m}(\zeta)d\zeta}{R(\zeta)} & \ldots & \oint_{\ggh}\frac{\zeta^{N-1}f_{\beta_k\beta_m}(\zeta)d\zeta}{R(\zeta)} &
\overline{\oint_{\ggh}\frac{f_{\beta_k\beta_m}(\zeta)d\zeta}{R(\zeta)}} & \ldots & \overline{\oint_{\ggh}\frac{\zeta^{N-1}f_{\beta_k\beta_m}(\zeta)d\zeta}{R(\zeta)}} &
\oint_{\ggh}\frac{f_{\beta_k\beta_m}(\zeta)d\zeta}{(\zeta-\alpha_j)R(\zeta)}
\end{array}\right|.
\end{equation}

Define $\vec K(\va,\vb)$ as
\[
\left\{\vec K(\va,\vb)\right\}_j = K(\alpha_j,\va,\vb), \ \ j=0,1,\ldots,2N+1.
\]

\begin{lemma}$ $\\
Let $f\in\mathbb{L}(\gamma_0,\va_0,\vb_0)$, where the contour $\gamma_0$ has the end points $\va_0$.
Then
\begin{equation}
K_j(\va,\vb):=K(\alpha_j,\va,\vb),\ \ \ \ j=0,1,\ldots, 2N+1,
\end{equation}
is continuously differentiable in $\va$ and in $\vb=(\beta_1,\ldots)$ in some open neighborhoods of $\va_0$ and $\vb_0$ respectively,
and for some contour $\gamma=\gamma(\va,\beta_1)$.
\label{lem:smooth_F}
\end{lemma}

\begin{proof}

$\vec{K}(\va,\bb)$ is analytic in $\va$ by the determinant structure and the integral entries (\ref{eq:K_function}),
where explicit dependence on $\va$ is only in the $R(z,\va)$ term which is analytic away from $z=\alpha_j$.

By Lemma \ref{lem:par_int_zero}, the partial derivatives $\frac{\partial K_j(\va,\vb)}{\partial \beta_k}$ exist and by twice differentiability
of $f(z,\bb)$ in $(\beta_2,\beta_3,\ldots)$ which leads to continuous $\frac{\partial^2 K(\alpha_j)}{\partial\beta_k\partial\beta_m}$, thus $\vec{K}(\va,\bb)$ is continuously differentiable in $(\beta_2,\beta_3,\ldots)$.

For $\beta_1$, the integrals in the last row of $\frac{\partial K(\alpha_j)}{\partial\beta_1}$ in (\ref{eq:DK/D_beta})
involve the function
\begin{equation}
f_{\beta_1}(z,\bb)=c_{\beta_1}(\bb)(z-z_0(\beta_1))\log(z-z_0(\beta_1))
+c(\bb)z'_0(\beta_1)\log(z-z_0(\beta_1))
-c(\bb)z'_0(\beta_1) +A_{\beta_1}(z,\bb),
\end{equation}
which is integrable near $z_0$ and hence $\frac{\partial K(\alpha_j)}{\partial\beta_1}$ is continuous in $\beta_1$.

To conclude joint smoothness of $K_j$ in $\bb$ notice that $K_j$ can be split into the
integrals of the analytic and the singular logarithmic parts of $f$ and notice that the
singular part of $f$, namely, $(z-z_0(\beta_1))\log(z-z_0(\beta_1))$ only depends on
$\beta_1$ as a function of $\bb$.
This allows to conclude the joint continuous differentiability of $K_j(\va,\vb)$ in $\bb$.

\end{proof}

Thus by Lemma \ref{lem:smooth_F} the modulation equations (\ref{K_eqn})
\begin{equation}
K_j(\vec\alpha,\vec\beta)=K(\alpha_j)=0
\label{eq:modulation}
\end{equation}
are smooth in $\va$ and in the external parameters $\vb$. Next we want to solve this system for $\va=\va(\vb)$
and conclude smoothness in $\vb$.

For the next lemma we need $K'(z,\va,\vb)=\frac{dK}{dz}(z,\va,\vb)$
\begin{equation}
K'(z,\va,\vb)=
\frac{1}{2\pi i}\left|\begin{array}{ccccccc}
\oint_{\ggh_{m,1}}\frac{d\zeta}{R(\zeta)} & \ldots & \oint_{\ggh_{m,1}}\frac{\zeta^{N-1}d\zeta}{R(\zeta)} &
\overline{\oint_{\ggh_{m,1}}\frac{d\zeta}{R(\zeta)}} & \ldots & \overline{\oint_{\ggh_{m,1}}\frac{\zeta^{N-1}d\zeta}{R(\zeta)}} & \oint_{\ggh_{m,1}}\frac{d\zeta}{(\zeta-z)^2R(\zeta)} \\
\ldots & \ldots & \ldots & \ldots & \ldots & \ldots & \ldots \\
\oint_{\ggh_{m,N}}\frac{d\zeta}{R(\zeta)} & \ldots & \oint_{\ggh_{m,N}}\frac{\zeta^{N-1}d\zeta}{R(\zeta)} &
\overline{\oint_{\ggh_{m,N}}\frac{d\zeta}{R(\zeta)}} & \ldots & \overline{\oint_{\ggh_{m,N}}\frac{\zeta^{N-1}d\zeta}{R(\zeta)}} &
\oint_{\ggh_{m,N}}\frac{d\zeta}{(\zeta-z)^2R(\zeta)} \\
\oint_{\ggh_{c,1}}\frac{d\zeta}{R(\zeta)} & \ldots & \oint_{\ggh_{c,1}}\frac{\zeta^{N-1}d\zeta}{R(\zeta)} &
\overline{\oint_{\ggh_{c,1}}\frac{d\zeta}{R(\zeta)}} & \ldots & \overline{\oint_{\ggh_{c,1}}\frac{\zeta^{N-1}d\zeta}{R(\zeta)}} &
\oint_{\ggh_{c,1}}\frac{d\zeta}{(\zeta-z)^2R(\zeta)} \\
\ldots & \ldots & \ldots & \ldots & \ldots & \ldots & \ldots \\
\oint_{\ggh_{c,N}}\frac{d\zeta}{R(\zeta)} & \ldots & \oint_{\ggh_{c,N}}\frac{\zeta^{N-1}d\zeta}{R(\zeta)} &
\overline{\oint_{\ggh_{c,N}}\frac{d\zeta}{R(\zeta)}} & \ldots & \overline{\oint_{\ggh_{c,N}}\frac{\zeta^{N-1}d\zeta}{R(\zeta)}} &
\oint_{\ggh_{c,N}}\frac{d\zeta}{(\zeta-z)^2R(\zeta)} \\
\oint_{\ggh}\frac{f(\zeta)d\zeta}{R(\zeta)} & \ldots & \oint_{\ggh}\frac{\zeta^{N-1}f(\zeta)d\zeta}{R(\zeta)} &
\overline{\oint_{\ggh}\frac{f(\zeta)d\zeta}{R(\zeta)}} & \ldots & \overline{\oint_{\ggh}\frac{\zeta^{N-1}f(\zeta)d\zeta}{R(\zeta)}} &
\oint_{\ggh}\frac{f(\zeta)d\zeta}{(\zeta-z)^2R(\zeta)}
\end{array}\right|,
\label{eq:K'_function}
\end{equation}
where $z$ is inside of $\ggh(\beta_1)$ and inside of $\ggh_{m,j}$ and $\ggh_{c,j}$ or $\ggh_{c,j+1}$.

\begin{lemma}$ $\\
Let $f\in\mathbb{L}(\gamma_0,\va_0,\vb_0)$, where $\va_0$ and $\vb_0$ satisfy
\[
\vec{K}(\va_0,\vb_0)=\vec{0}.
\]
Assume that for $\va_0=\left\{\alpha^{0}_j\right\}_{j=0}^{2N+1}$,
$\lim_{z\to\alpha^0_j}K'(z,\va_0,\vb_0)\neq 0,$ $j=0, 1,\ldots, 2N+1$.

Then the modulation equations
\[
\vec{K}(\va,\vb)=\vec{0}
\]
can be uniquely solved for $\va(\vb)$ which is continuously
differentiable for all $\vb$ in some open neighborhood of $\vb_0$ and $\va(\vb_0)=\va_0$.
\label{lem:smooth_alpha}
\end{lemma}

\begin{proof}

$\vec{K}$ is continuously differentiable in $\va$ and $\vec{K}$ is a continuously differentiable in
 $\vb$ by Lemma \ref{lem:smooth_F}.

As it was shown in \cite{TV_det}, the matrix
\begin{equation}
\left\{\frac{\partial \vec{K}}{\partial\va}\right\}_{j,l}
=\left\{\frac{\partial K(\alpha_j)}{\partial\alpha_l}\right\}_{j,l}
\end{equation}
is diagonal and
\begin{equation}
\frac{\partial K(\alpha_j)}{\partial\alpha_j}=\frac{3}{2}D\lim_{z\to\alpha^0_j}\left(\frac{h(z)}{R(z)}\right)'=
\frac{3}{2}\lim_{z\to\alpha^0_j}K'(z,\va,\vb) \neq 0.
\end{equation}
So
\begin{equation}
\det\left|\frac{\partial \vec{K}}{\partial\vec\alpha}(\va_0)\right|=\prod_{j}\frac{\partial K(\alpha_j)}{\partial\alpha_j}\neq 0
\end{equation}
under the assumptions. By the Implicit function theorem, $\va(\vb)$
are uniquely defined in some neighborhood of $\vb_0$ and smooth in $\vb$.
Note $\va(\vb_0)=\va_0$ by assumption.
\end{proof}

\begin{remark}$ $\\
The condition $\lim_{z\to\alpha^{0}_j}K'(z,\va_0,\vb_0)\neq 0,$ $j=0, 1,\ldots, 2N+1$ in Lemma
\ref{lem:smooth_alpha} is equivalent to $\lim_{z\to\alpha^{0}_j}\frac{h'(z,\va_0,\vb_0)}{R(z,\va_0)}\neq 0,$ $j=0, 1,\ldots, 2N+1$.

\end{remark}

To summarize this section, given the set of $\va_0$, $\vb_0=(\beta^0_1,...)$, $\gamma_0=\gamma(\va_0,\beta^0_1)$, and $f\in\mathbb{L}(\gamma_0,\va_0,\vb_0)$, defines
the function $K(z,\va_0,\vb_0)$ by formula (\ref{eq:K_function}) which through (\ref{eq:h_and_K}) in turn defines $h(z,\va_0,\vb_0)$,
which satisfies a scalar Riemann-Hilbert problem on $\gamma_0$:
\begin{equation}
\left\{
\begin{array}{l}
h_{+}(z)+h_{-}(z)=2W_j(\va_0,\vb_0),\ \mbox{on} \ \gg_{m,j}, \ \ j=0,1,...,N,\\
h_{+}(z)-h_{-}(z)=2\Omega_j(\va_0,\vb_0),\ \mbox{on} \ \gg_{c,j}, \ \ j=1,...,N,\\
h(z)+f(z)\ \mbox{is analytic in}\ \mathbb{\overline{C}}\backslash\gg_0.
\end{array}
\right.
\label{eq:RHP_h_at_vb_0}
\end{equation}
We denote this scalar Riemann-Hilbert problem as $RHP(\gamma_0,\va_0,\vb_0,f)$.

Lemma \ref{lem:smooth_alpha} states that if $\va_0$ and $\vb_0$ satisfy $\vec{K}(\va_0,\vb_0)=\vec{0}$, and $RHP(\gamma_0,\va_0,\vb_0,f)$ has a
solution with $\lim_{z\to\alpha^0_j}K'(z,\va_0,\vb_0)\neq 0$ then $RHP(\gamma,\va,\vb,f)$
has a unique smooth solution for all $\va=\va(\vb)$ and $\vb$ in some open neighborhood of $\vb_0$,
where $\va$ and $\vb$ solve the modulation equations $\vec{K}(\va,\vb)=\vec{0}$.

\section{Perturbation theorem}

\begin{theorem} (Perturbation Theorem)\\
Consider a simple contour $\gamma_0$ consisting of a finite union of oriented simple arcs $\gamma_0=\left(\bigcup\gg_{m,j}\right)\cup\left(\bigcup\gg_{c,j}\right)$
 with the distinct end points $\va_0$ and depending on parameters $\vb_0$ (see Figure \ref{fig:RHP_contours}). Assume $\va_0$ and $\vb_0$
satisfy a system of equations
\[
\vec{K}(\va_0,\vb_0)=\vec{0},
\label{eq:K_0}
\]
and let $f\in\mathbb{L}(\gamma,\va_0,\vb_0)$.
Let $\gamma=\gamma(\va,\beta_1)$ be the contour of a RH problem which seeks a function
$h(z)=h(z,\va,\vb)$ which satisfies the following conditions
\begin{equation}
\left\{
\begin{array}{l}
h_{+}(z)+h_{-}(z)=2W_j,\ \mbox{on} \ \gg_{m,j}, \ \ j=0,1,...,N,\\
h_{+}(z)-h_{-}(z)=2\Omega_j,\ \mbox{on} \ \gg_{c,j}, \ \ j=1,...,N,\\
h(z)+f(z)\ \mbox{is analytic in}\ \mathbb{\overline{C}}\backslash\gg,
\end{array}
\right. \label{eq:RHP_h}
\end{equation}
where $\Omega_j=\Omega_j(\va,\vb)$ and $W_j=W_j(\va,\vb)$ are real constants whose numerical values will be determined from the RH conditions.
Assume that there is a function $h(z,\va_0,\vb_0)$ which satisfies (\ref{eq:RHP_h}) and suppose $\frac{h'(z,\va_0,\vb_0)}{R(z,\va_0)}\neq 0$
for all $z$ on $\gamma_0$.

Then the solutions $\va=\va(\vb)$ of the system
\begin{equation}
\vec{K}(\va,\vb)=\vec{0}
\label{eq:vec_K_system}
\end{equation}
and $h(z,\va(\vb),\vb)$ which solves (\ref{eq:RHP_h}) are uniquely defined and smooth in some open neighborhood of $\vb_0$.
Moreover, $\Omega_j(\vb)=\Omega_j(\va(\vb),\vb)$, and $W_j(\vb)=W_j(\va(\vb),\vb)$ are defined and smooth in $\bb$ in some open neighborhood of $\bb_0$.

Furthermore, for $k\ge 1$:
\begin{equation}
\frac{\partial\alpha_j}{\partial\beta_k}(\vb)=-\frac{2\pi i \frac{\partial K(\alpha_j,\va(\vb),\bb)}{\partial\beta_k}}{D(\va(\vb),\vb) \oint_{\ggh(\beta_1)}\frac{f'(\zeta,\vb)}{(\zeta-\alpha_j(\vb))R(\zeta,\va(\vb))}d\zeta}, \ \ j=1,2,...,2N+1,
\label{eq:alpha_mu}
\end{equation}
\begin{equation}
\frac{\partial h}{\partial\beta_k}(z,\va,\vb)=\frac{R(z,\va)}{2\pi i}\int_{\ggh(\beta_1)}
\frac{\frac{\partial f}{\partial\beta_k}(\xi,\vb)}{(\xi-z)R(\xi,\va)}d\xi,
\label{eq:h_mu}
\end{equation}
where $z$ is inside of $\ggh(\beta_1)$,
\begin{equation}
\frac{\partial\Omega_j}{\partial\b_k}(\vb)=\frac{-1}{D}
\left|\begin{array}{cccccc}
\oint_{\ggh_{m,1}}\frac{d\zeta}{R(\zeta)} & \ldots & \oint_{\ggh_{m,1}}\frac{\zeta^{N-1}d\zeta}{R(\zeta)} &
\overline{\oint_{\ggh_{m,1}}\frac{d\zeta}{R(\zeta)}} & \ldots & \overline{\oint_{\ggh_{m,1}}\frac{\zeta^{N-1}d\zeta}{R(\zeta)}} \\
\ldots & \ldots & \ldots & \ldots & \ldots & \ldots \\
\oint_{\ggh_{m,N}}\frac{d\zeta}{R(\zeta)} & \ldots & \oint_{\ggh_{m,N}}\frac{\zeta^{N-1}d\zeta}{R(\zeta)} &
\overline{\oint_{\ggh_{m,N}}\frac{d\zeta}{R(\zeta)}} & \ldots & \overline{\oint_{\ggh_{m,N}}\frac{\zeta^{N-1}d\zeta}{R(\zeta)}}  \\
\oint_{\ggh_{c,1}}\frac{d\zeta}{R(\zeta)} & \ldots & \oint_{\ggh_{c,1}}\frac{\zeta^{N-1}d\zeta}{R(\zeta)} &
\overline{\oint_{\ggh_{c,1}}\frac{d\zeta}{R(\zeta)}} & \ldots & \overline{\oint_{\ggh_{c,1}}\frac{\zeta^{N-1}d\zeta}{R(\zeta)}} \\
\ldots & \ldots & \ldots & \ldots & \ldots & \ldots  \\
\oint_{\ggh_{c,j-1}}\frac{d\zeta}{R(\zeta)} & \ldots & \oint_{\ggh_{c,j-1}}\frac{\zeta^{N-1}d\zeta}{R(\zeta)} &
\overline{\oint_{\ggh_{c,j-1}}\frac{d\zeta}{R(\zeta)}} & \ldots & \overline{\oint_{\ggh_{c,j-1}}\frac{\zeta^{N-1}d\zeta}{R(\zeta)}} \\
{\oint_{\ggh(\beta_1)}\frac{ f_{\beta_k}(\xi,\bb)}{R(\xi,\va)}d\xi} & \ldots & {\oint_{\ggh(\beta_1)}\frac{\xi^{N-1} f_{\beta_k}(\xi,\bb)}{R(\xi,\va)}d\xi}& \overline{\oint_{\ggh(\beta_1)}\frac{ f_{\beta_k}(\xi,\bb)}{R(\xi,\va)}d\xi} & \ldots &
\overline{\oint_{\ggh(\beta_1)}\frac{\xi^{N-1} f_{\beta_k}(\xi,\bb)}{R(\xi,\va)}d\xi}\\
\oint_{\ggh_{c,j+1}}\frac{d\zeta}{R(\zeta)} & \ldots & \oint_{\ggh_{c,j+1}}\frac{\zeta^{N-1}d\zeta}{R(\zeta)} &
\overline{\oint_{\ggh_{c,j+1}}\frac{d\zeta}{R(\zeta)}} & \ldots & \overline{\oint_{\ggh_{c,j+1}}\frac{\zeta^{N-1}d\zeta}{R(\zeta)}} \\
\ldots & \ldots & \ldots & \ldots & \ldots & \ldots  \\
\oint_{\ggh_{c,N}}\frac{d\zeta}{R(\zeta)} & \ldots & \oint_{\ggh_{c,N}}\frac{\zeta^{N-1}d\zeta}{R(\zeta)} &
\overline{\oint_{\ggh_{c,N}}\frac{d\zeta}{R(\zeta)}} & \ldots & \overline{\oint_{\ggh_{c,N}}\frac{\zeta^{N-1}d\zeta}{R(\zeta)}}
\end{array}\right|,
\label{eq:Omega_mu}
\end{equation}
\begin{equation}
\frac{\partial W_j}{\partial\beta_k}(\vb)=\frac{-1}{D}
\left|\begin{array}{cccccc}
\oint_{\ggh_{m,1}}\frac{d\zeta}{R(\zeta)} & \ldots & \oint_{\ggh_{m,1}}\frac{\zeta^{N-1}d\zeta}{R(\zeta)} &
\overline{\oint_{\ggh_{m,1}}\frac{d\zeta}{R(\zeta)}} & \ldots & \overline{\oint_{\ggh_{m,1}}\frac{\zeta^{N-1}d\zeta}{R(\zeta)}} \\
\ldots & \ldots & \ldots & \ldots & \ldots & \ldots \\
\oint_{\ggh_{m,j-1}}\frac{d\zeta}{R(\zeta)} & \ldots & \oint_{\ggh_{m,j-1}}\frac{\zeta^{N-1}d\zeta}{R(\zeta)} &
\overline{\oint_{\ggh_{m,j-1}}\frac{d\zeta}{R(\zeta)}} & \ldots & \overline{\oint_{\ggh_{m,j-1}}\frac{\zeta^{N-1}d\zeta}{R(\zeta)}} \\
{\oint_{\ggh(\beta_1)}\frac{ f_{\beta_k}(\xi,\bb)}{R(\xi,\va)}d\xi} & \ldots & {\oint_{\ggh(\beta_1)}\frac{\xi^{N-1} f_{\beta_k}(\xi,\bb)}{R(\xi,\va)}d\xi}& \overline{\oint_{\ggh(\beta_1)}\frac{ f_{\beta_k}(\xi,\bb)}{R(\xi,\va)}d\xi} & \ldots &
\overline{\oint_{\ggh(\beta_1)}\frac{\xi^{N-1} f_{\beta_k}(\xi,\bb)}{R(\xi,\va)}d\xi}\\
\oint_{\ggh_{m,j+1}}\frac{d\zeta}{R(\zeta)} & \ldots & \oint_{\ggh_{m,j+1}}\frac{\zeta^{N-1}d\zeta}{R(\zeta)} &
\overline{\oint_{\ggh_{m,j+1}}\frac{d\zeta}{R(\zeta)}} & \ldots & \overline{\oint_{\ggh_{m,j+1}}\frac{\zeta^{N-1}d\zeta}{R(\zeta)}} \\
\ldots & \ldots & \ldots & \ldots & \ldots & \ldots  \\
\oint_{\ggh_{m,N}}\frac{d\zeta}{R(\zeta)} & \ldots & \oint_{\ggh_{m,N}}\frac{\zeta^{N-1}d\zeta}{R(\zeta)} &
\overline{\oint_{\ggh_{m,N}}\frac{d\zeta}{R(\zeta)}} & \ldots & \overline{\oint_{\ggh_{m,N}}\frac{\zeta^{N-1}d\zeta}{R(\zeta)}}  \\
\oint_{\ggh_{c,1}}\frac{d\zeta}{R(\zeta)} & \ldots & \oint_{\ggh_{c,1}}\frac{\zeta^{N-1}d\zeta}{R(\zeta)} &
\overline{\oint_{\ggh_{c,1}}\frac{d\zeta}{R(\zeta)}} & \ldots & \overline{\oint_{\ggh_{c,1}}\frac{\zeta^{N-1}d\zeta}{R(\zeta)}} \\
\ldots & \ldots & \ldots & \ldots & \ldots & \ldots  \\
\oint_{\ggh_{c,N}}\frac{d\zeta}{R(\zeta)} & \ldots & \oint_{\ggh_{c,N}}\frac{\zeta^{N-1}d\zeta}{R(\zeta)} &
\overline{\oint_{\ggh_{c,N}}\frac{d\zeta}{R(\zeta)}} & \ldots & \overline{\oint_{\ggh_{c,N}}\frac{\zeta^{N-1}d\zeta}{R(\zeta)}}
\end{array}\right|,
\label{eq:W_mu}
\end{equation}
where $R(\xi)=R(\xi,\va(\vb))$.

\label{thm:perturb}
\end{theorem}

\begin{remark}
Formula (\ref{eq:alpha_mu}) is the same as (56) in \cite{TV_det}. We prove that the formula is still valid when
the jump contour $\gamma$ depends on an external parameter ($\beta_1$ in our case). The contour depends
on $\beta_1$ through the logarithmic singularity $z_0(\beta_1)$ on $\gamma$ (see the paragraph
after the definition \ref{def:3.1}).

\end{remark}

\begin{proof}

By Lemma \ref{lem:smooth_alpha} $\alpha_j(\vb)$ are continuously differentiable in $\vb$. Formula for $\frac{\partial\alpha_j}{\partial\beta_k}$
were computed above (\ref{eq_alpha_mu}).

Next we compute $\frac{\partial g(z,\vec\alpha,\bb)}{\partial\beta_k}$ which satisfies the RHP
\begin{equation}
g_{\beta_k,+}(z)-g_{\beta_k,-}(z)=f_{\beta_k}(z),\ \ \ z\in \gamma_{m,j}, \ j=0,1,...,N.
\end{equation}
Then
\begin{equation}
\frac{\partial g}{\partial\beta_k}(z,\va,\vb)=\frac{R(z,\va)}{2\pi i}\int_{\ggh}
\frac{\frac{\partial f}{\partial\beta_k}(\xi,\vb)}{(\xi-z)R(\xi,\va)}d\xi
\end{equation}
where $z$ is outside of $\ggh$, $\frac{\partial f}{\partial\beta_k}(z)$ for $k=1$ behaves like $\log(z-z_0)$ near $z_0$,
and for $k\ge 2$ behaves like $(z-z_0)\log(z-z_0)$ near $z_0$.
So $\frac{\partial h}{\partial\beta_k}(z,\va,\vb)$ satisfies (\ref{eq:h_mu}) where $z$ is inside of $\ggh$.

Constants $W_j$ and $\Omega_j$ are found from the linear system \cite{TV_det}
\begin{equation}
\oint_{\ggh(\beta_1)}\frac{\xi^n f(\xi,\bb)}{R(\xi,\va)}d\xi +
\sum_{j=1}^{N}\oint_{\ggh_{c,j}}\frac{\xi^n \Omega_j}{R(\xi,\va)}d\xi+
\sum_{j=1}^{N}\oint_{\ggh_{m,j}(\beta_1)}\frac{\xi^n W_j}{R(\xi,\va)}d\xi =0, n=0,...N-1.
\end{equation}

Differentiating in $\beta_k$ and using Lemma \ref{lem:par_int_zero} leads to

\begin{equation}
\oint_{\ggh(\beta_1)}\frac{\xi^n f_{\beta_k}(\xi,\bb)}{R(\xi,\va)}d\xi +
\sum_{j=1}^{N}\oint_{\ggh_{c,j}}\frac{\xi^n \Omega_{j,\beta_k}}{R(\xi,\va)}d\xi+
\sum_{j=1}^{N}\oint_{\ggh_{m,j}(\beta_1)}\frac{\xi^n W_{j,\beta_k}}{R(\xi,\va)}d\xi =0, n=0,...,N-1
\end{equation}
and since by assumption $W_j$ and $\Omega_j$ are real and they satisfy the complex conjugate system as well
\begin{equation}
\overline{\oint_{\ggh(\beta_1)}\frac{\xi^n f_{\beta_k}(\xi,\bb)}{R(\xi,\va)}d\xi} +
\overline{\sum_{j=1}^{N}\oint_{\ggh_{c,j}}\frac{\xi^n \Omega_{j,\beta_k}}{R(\xi,\va)}d\xi}+
\overline{\sum_{j=1}^{N}\oint_{\ggh_{m,j}(\beta_1)}\frac{\xi^n W_{j,\beta_k}}{R(\xi,\va)}d\xi} =0, n=0,...,N-1,
\end{equation}
or in matrix form
\begin{equation}
\left(\begin{array}{cccccc}
\oint_{\ggh_{m,1}}\frac{d\zeta}{R(\zeta)} & \ldots & \oint_{\ggh_{m,1}}\frac{\zeta^{N-1}d\zeta}{R(\zeta)} &
\overline{\oint_{\ggh_{m,1}}\frac{d\zeta}{R(\zeta)}} & \ldots & \overline{\oint_{\ggh_{m,1}}\frac{\zeta^{N-1}d\zeta}{R(\zeta)}} \\
\ldots & \ldots & \ldots & \ldots & \ldots & \ldots \\
\oint_{\ggh_{m,N}}\frac{d\zeta}{R(\zeta)} & \ldots & \oint_{\ggh_{m,N}}\frac{\zeta^{N-1}d\zeta}{R(\zeta)} &
\overline{\oint_{\ggh_{m,N}}\frac{d\zeta}{R(\zeta)}} & \ldots & \overline{\oint_{\ggh_{m,N}}\frac{\zeta^{N-1}d\zeta}{R(\zeta)}}  \\
\oint_{\ggh_{c,1}}\frac{d\zeta}{R(\zeta)} & \ldots & \oint_{\ggh_{c,1}}\frac{\zeta^{N-1}d\zeta}{R(\zeta)} &
\overline{\oint_{\ggh_{c,1}}\frac{d\zeta}{R(\zeta)}} & \ldots & \overline{\oint_{\ggh_{c,1}}\frac{\zeta^{N-1}d\zeta}{R(\zeta)}} \\
\ldots & \ldots & \ldots & \ldots & \ldots & \ldots  \\
\oint_{\ggh_{c,N}}\frac{d\zeta}{R(\zeta)} & \ldots & \oint_{\ggh_{c,N}}\frac{\zeta^{N-1}d\zeta}{R(\zeta)} &
\overline{\oint_{\ggh_{c,N}}\frac{d\zeta}{R(\zeta)}} & \ldots & \overline{\oint_{\ggh_{c,N}}\frac{\zeta^{N-1}d\zeta}{R(\zeta)}}
\end{array}\right)^T
\left(\begin{array}{c}
\frac{\partial\vec W}{\partial\b_k}\\
\frac{\partial\vec\Omega}{\partial\b_k}
\end{array}\right)
=-\left(\begin{array}{c}
{\oint_{\ggh(\beta_1)}\frac{f_{\beta_k}(\xi,\bb)}{R(\xi,\va)}d\xi}\\
\ldots\\
{\oint_{\ggh(\beta_1)}\frac{\xi^{N-1} f_{\beta_k}(\xi,\bb)}{R(\xi,\va)}d\xi}\\
\overline{\oint_{\ggh(\beta_1)}\frac{f_{\beta_k}(\xi,\bb)}{R(\xi,\va)}d\xi}\\
\ldots\\
\overline{\oint_{\ggh(\beta_1)}\frac{\xi^{N-1} f_{\beta_k}(\xi,\bb)}{R(\xi,\va)}d\xi}
\end{array}\right).
\end{equation}
So $\frac{\partial\Omega_j}{\partial\b_k}$ and $\frac{\partial W_j}{\partial\beta_k}$ satisfy (\ref{eq:Omega_mu}) and (\ref{eq:W_mu}).
Note that $D\neq 0$ for distinct $\alpha_j$'s \cite{TV_det}.

\end{proof}

\begin{figure}
\begin{center}
\includegraphics[height=5cm]{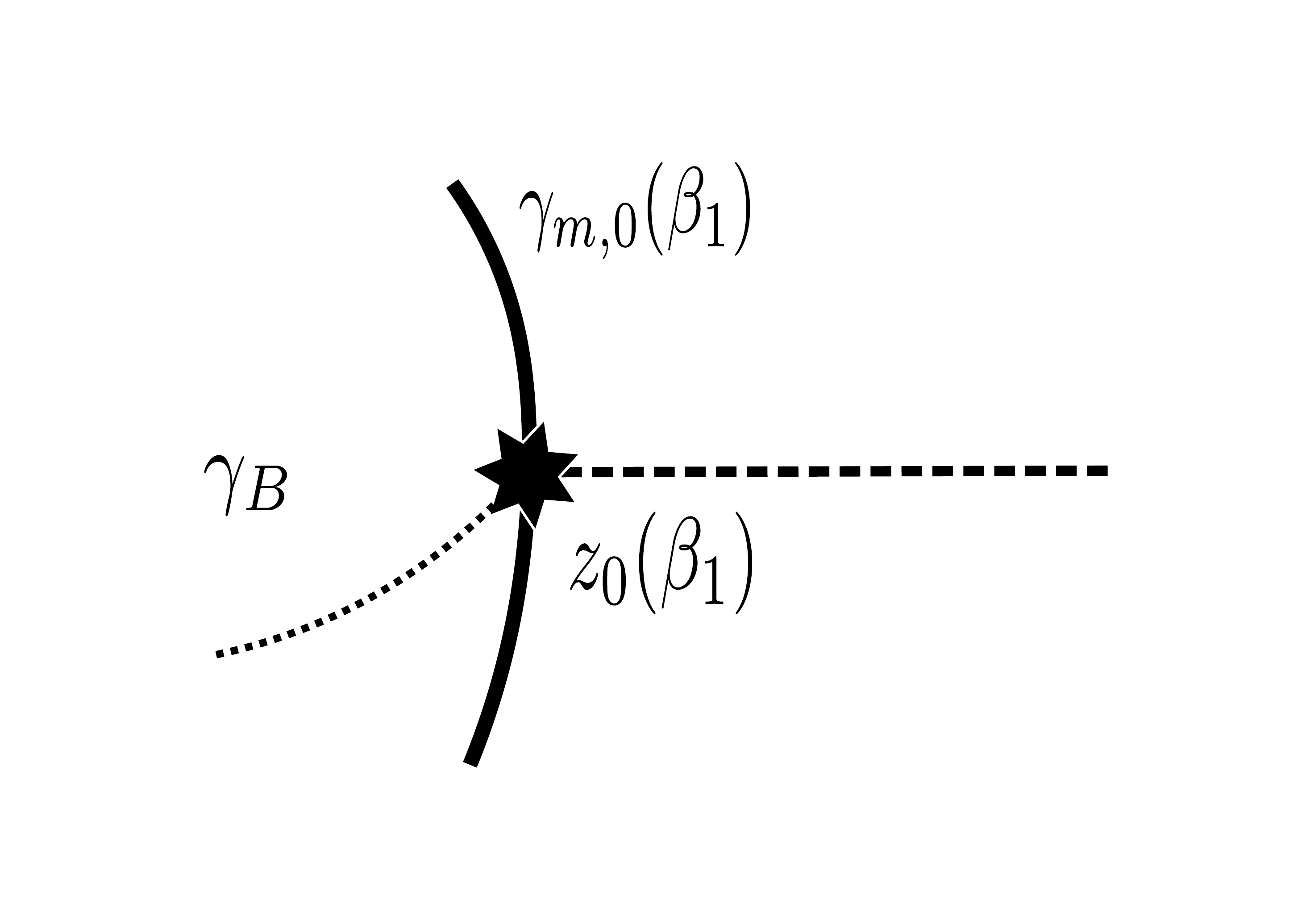}
\caption{\label{fig:new_f_jump_contour} The dashed curves are the jump contours
of $B$ and $f$ near $z_0$ in definition \ref{def:L_tilde}.}
\end{center}
\end{figure}

To apply this theorem to the semiclassical focusing NLS we make a slight modification of the class
of the allowed functions in the definition \ref{def:3.1}. In the case of NLS, the real axis is an
additional contour of discontinuity of $f$ because of the Schwarz symmetry condition. This additional
jump contour of $f$ affects only a small neighborhood of $z_0$.
\begin{definition}
We say that $f\in\mathbb{\tilde{L}}(\gamma_0,\va_0,\vb_0)$ if
$f\in\mathbb{L}(\gamma_0,\va_0,\vb_0)$ with the condition (\ref{eq:f_form_def}) is
replaced with
\begin{equation}
\left\{\begin{array}{l}
f(z,\vb)=c(\vb)\left(z-z_0\right)\log\left(z-z_0\right)+(z-z_0)B(z,\vb)+A(z,\vb),\\
\lim_{z\to z_0,\ z\in\gamma_0}f(z,\vb)\ \ \mbox{exists and finite},
\end{array}\right.
\label{eq:f_form_def_tilde}
\end{equation}
where for all $\vb$ in some open neighborhood of $\vb_0$, $B(z,\vb)$ has a constant jump
in $z$ near $z_0$ on a simple contour $\gamma_B$ intersecting $\gamma_0$ only at $z_0$
and coinciding with the branchcut of $\log\left(z-z_0\right)$ near $z_0$.
Assume $c(\vb)$ and $A(z,\vb)$ satisfy the same conditions as before,
and $B(z,\vb)$ is twice continuously differentiable in parameters $\vb$ in some
open neighborhood of $\vb_0$;
and for each $\vb$ in some open neighborhood of $\vb_0$, $B(z,\vb)$ is piecewise
analytic in $z$ in some
open neighborhood of $\gg_0\backslash\{z_0\}$ in $\mathbb{C}\backslash\gg_B$.
\label{def:L_tilde}
\end{definition}
This technical modification changes the jump of $f$ on $[\delta_1,\delta_2]$
near $z_0$ in the proof of Lemma \ref{lem:par_int_zero}, however the jump of $f$ remains
linear and by the second condition in (\ref{eq:f_form_def_tilde}) the jump of $f$
is $0$ at $z_0$. This automatically is correct for $f\in\mathbb{L}(\gamma_0,\va_0,\vb_0)$
because of the $z\log z$ type singularity at $z_0$.
Thus for $f\in\mathbb{\tilde{L}}(\gamma,\va_0,\vb_0)$,
(\ref{eq:estimating_O(beta_1)_integral}-\ref{eq:I_1_with_O(beta_1)}) are correct and
all the above results including Lemma \ref{lem:f_in_L_near}, Lemma \ref{lem:smooth_alpha} and Theorem \ref{thm:perturb}
remain valid.

This definition was not used from the beginning since the main difficulty
is to deal with the jump of logarithm in (\ref{eq:f_form_def_tilde}) while
the jump of $B$ is a technical issue which would make the exposition
less clear.

\section{$\mu$-dependence in the semiclassical focusing NLS}

The function $f(z)$ comes from a semiclassical approximation of the family of initial
conditions for NLS (\ref{our_IC_NLS}) as \cite{TVZzero}:
\[
f(z,\mu,x,t)=\left(\frac{\mu}{2}-z\right)\left[\frac{\pi
i}{2}+\ln\left(\frac{\mu}{2}-z\right)\right]+\frac{z+T}{2}\ln\left(z+T\right)
+\frac{z-T}{2}\ln\left(z-T\right)
\]
\begin{equation}
-T\tanh^{-1}\frac{T}{\mu/2}-xz-2tz^2 +\frac{\mu}{2}\ln 2, \quad
\mbox{when}\ \ \Im z\ge 0 \label{eq:general f-function}
\end{equation}
and
\begin{equation}
f(z)=\overline{f(\overline{z})},\quad \mbox{when}\ \ \Im z< 0,
\label{eq:Schwarz_reflection}
\end{equation}
where the branchcuts are chosen as the following: from $\mm$ along the real
axis to $+\infty$, from $T$ to $0$ and along the real axis to $+\infty$, from
$-T$ to $0$ and along the real axis to $-\infty$, where
\begin{equation}
T=\sqrt{\frac{\mu^2}{4}-1},\ \ \ \Im T\ge 0.
\end{equation}
For $\mu\ge 2$, $T\ge 0$ is real and for $0<\mu<2$, $T$ is purely imaginary with $\Im T>0$.

Then for $\Im z\ge 0$
\begin{equation}
\frac{\partial f}{\partial\mu}(z,\mu)=\frac{\pi i}{4}+\frac{1}{2}\ln\left(\frac{\mu}{2}-z\right)+\ln 2+ \frac{\mu}{8T}\left[\ln(z+T)-\ln(z-T)-2\tanh^{-1}\frac{2T}{\mu}\right]
\label{eq_f'_mu}
\end{equation}

where $\tanh^{-1} x = x+O(x^3),\ x\to 0$ then for $z\neq0$

\begin{equation}
\frac{\partial f}{\partial\mu}(z,\mu) = \frac{\pi i}{4}+\frac{1}{2}\ln\left(\frac{\mu}{2}-z\right)+\ln 2+ \frac{\mu}{4z} -\frac{1}{2}+O(T),\ T\to 0.
\end{equation}

So $\mu=2$ is a removable singularity for $f_\mu$ as a function of $\mu$
\begin{equation}
\lim_{\mu\to 2, T\to 0} \frac{\partial f}{\partial\mu}(z,\mu)=\frac{\pi i}{4}+\frac{1}{2}\ln\left(1-z\right)+\ln 2 + \frac{1}{2z} -\frac{1}{2},
\end{equation}
which is analytic for $z\neq0$ and $z\neq1$.

\begin{lemma}
$f(z,\mu,x,t)$ and $f'(z,\mu,x,t)$ are analytic in $\mu$ for $\mu>0$, $x>0$, $t>0$, for all $z$, $\Im z\neq 0$.  
\label{lem:f_analytic}
\end{lemma}

\begin{proof}

Consider
\begin{equation}
f'(z,\mu,x,t)=-\frac{\pi i}{2}-\ln\left(\frac{\mu}{2}-z\right)
+\frac{1}{2}\ln\left(z^2-\frac{\mu^2}{4}+1\right)-x-4tz,
\end{equation}
which analytic in $\mu>0$, for $\Im z\neq 0$. 

For $\mu>0$, $\mu\neq 2$, $f(z,\mu)$ is clearly analytic in $\mu$ for $\Im z\neq 0$. At $\mu=2$ ($T=0$) we find the power
series of $f(z,\mu,x,t)$ in $T$ and show that it contains only even powers. Since
\begin{equation}
T^{2k}=\left(\frac{\mu^2}{4}-1\right)^{k}=\frac{(\mu+2)^k}{4^k}(\mu-2)^k
\end{equation}
it will show analyticity of $f(z,\mu,x,t)$ in $\mu$.

Start with expanding in series at $T=0$
\begin{equation}
\frac{1}{\mu/2}=\sqrt{1+T^2}^{-1}=\sum_{k=0}^\infty c_k T^{2k},\ \ \ \ln\left(z\pm T\right)=\sum_{n=1}^{\infty}\frac{(-1)^{n+1}}{n}\left(\pm\frac{T}{z}\right)^n.
\end{equation}
Then the terms in (\ref{eq:general f-function}) are analytic in $\mu$
\begin{equation}
\frac{z+T}{2}\ln\left(z+T\right)+\frac{z-T}{2}\ln\left(z-T\right)
\end{equation}
\begin{equation}
=z\ln z - z \sum_{n\ is\ even} \frac{1}{n}\left(\frac{T}{z}\right)^n + T\sum_{n\ is\ odd}\frac{1}{n}\left(\frac{T}{z}\right)^n
\end{equation}
\begin{equation}
=z\ln z + \sum_{k=1}^{\infty} \frac{1}{2k(2k-1)z^{2k-1}} T^{2k},
\end{equation}
which has only even powers of $T$ and is analytic in $\mu$ for $\Im z\neq 0$.
Consider the inverse hyperbolic tangent term in (\ref{eq:general f-function}) and
take into account that $\tanh^{-1}z$ is an odd function
\begin{equation}
T\tanh^{-1}\frac{T}{\mu/2}=T\tanh^{-1}\frac{T}{\sqrt{1+T^2}}
=T\tanh^{-1}\sum_{k=0}^\infty c_k T^{2k+1}
\end{equation}
\begin{equation}
=T\sum_{k=0}^\infty \tilde{c}_k T^{2k+1}=\sum_{k=0}^\infty \tilde{c}_k T^{2k+2},
\end{equation}
which also has only even powers of $T$.

So
\[
f(z)=\left(\frac{\mu}{2}-z\right)\left[\frac{\pi
i}{2}+\ln\left(\frac{\mu}{2}-z\right)\right]+\frac{z+T}{2}\ln\left(z+T\right)
+\frac{z-T}{2}\ln\left(z-T\right)
\]
\begin{equation}
-T\tanh^{-1}\frac{T}{\mu/2}-xz-2tz^2 +\frac{\mu}{2}\ln 2
\end{equation}
is analytic in $\mu$ for $\mu>0$, $x>0$, $t>0$, $\Im z\neq 0$.

\end{proof}

\begin{lemma}$ $\\
Let $f(z)$ be given by (\ref{eq:general f-function}) and the contour $\gamma_0$ consists of the union of oriented simple arcs $\gamma_0=\left(\bigcup\gg_{m,j}\right)\cup\left(\bigcup\gg_{c,j}\right)$ with distinct end points $\vec\alpha_0$. Assume
that $\gamma_0$ only crosses the real axis at $z_0=\frac{\mu}{2}$ and $\gamma_0$ does not cross the interval $[-T,T]$. Then
$f\in\mathbb{\tilde{L}}(\gamma,\va,\vb)$, where $\vb=(\mu,x,t)$, for all $x>0$, $t>0$, and $\mu>0$.
\label{lem:f_in_L_NLS}
\end{lemma}

\begin{proof}
Follows from Definition \ref{def:L_tilde} with the logarithmic singularity at
$z_0=\frac{\mu}{2}$, Lemma \ref{lem:f_analytic}, and Lemma \ref{lem:f_in_L_near}.

Note the jump of $f(z)$ is caused by the Schwarz reflection (\ref{eq:Schwarz_reflection}) on the real axis and
it is linear in $z$ since $\Im f$ is a linear function on the real axis (as a limit)
near $\mm$ with $\Im f \left(\mm\right)=0$ \cite{TVZzero}.

\end{proof}

Now we can apply all results from Section 4 including the Perturbation theorem \ref{thm:perturb}.

The main objects are
\begin{equation}
K(\alpha_j,\va,\mu,x,t)=
\frac{1}{2\pi i}\left|\begin{array}{ccccccc}
\oint_{\ggh_{m,1}}\frac{d\zeta}{R(\zeta)} & \ldots & \oint_{\ggh_{m,1}}\frac{\zeta^{N-1}d\zeta}{R(\zeta)} &
\overline{\oint_{\ggh_{m,1}}\frac{d\zeta}{R(\zeta)}} & \ldots & \overline{\oint_{\ggh_{m,1}}\frac{\zeta^{N-1}d\zeta}{R(\zeta)}} & \oint_{\ggh_{m,1}}\frac{d\zeta}{(\zeta-\alpha_j)R(\zeta)} \\
\ldots & \ldots & \ldots & \ldots & \ldots & \ldots & \ldots \\
\oint_{\ggh_{m,N}}\frac{d\zeta}{R(\zeta)} & \ldots & \oint_{\ggh_{m,N}}\frac{\zeta^{N-1}d\zeta}{R(\zeta)} &
\overline{\oint_{\ggh_{m,N}}\frac{d\zeta}{R(\zeta)}} & \ldots & \overline{\oint_{\ggh_{m,N}}\frac{\zeta^{N-1}d\zeta}{R(\zeta)}} &
\oint_{\ggh_{m,N}}\frac{d\zeta}{(\zeta-\alpha_j)R(\zeta)} \\
\oint_{\ggh_{c,1}}\frac{d\zeta}{R(\zeta)} & \ldots & \oint_{\ggh_{c,1}}\frac{\zeta^{N-1}d\zeta}{R(\zeta)} &
\overline{\oint_{\ggh_{c,1}}\frac{d\zeta}{R(\zeta)}} & \ldots & \overline{\oint_{\ggh_{c,1}}\frac{\zeta^{N-1}d\zeta}{R(\zeta)}} &
\oint_{\ggh_{c,1}}\frac{d\zeta}{(\zeta-\alpha_j)R(\zeta)} \\
\ldots & \ldots & \ldots & \ldots & \ldots & \ldots & \ldots \\
\oint_{\ggh_{c,N}}\frac{d\zeta}{R(\zeta)} & \ldots & \oint_{\ggh_{c,N}}\frac{\zeta^{N-1}d\zeta}{R(\zeta)} &
\overline{\oint_{\ggh_{c,N}}\frac{d\zeta}{R(\zeta)}} & \ldots & \overline{\oint_{\ggh_{c,N}}\frac{\zeta^{N-1}d\zeta}{R(\zeta)}} &
\oint_{\ggh_{c,N}}\frac{d\zeta}{(\zeta-\alpha_j)R(\zeta)} \\
\oint_{\ggh}\frac{f(\zeta)d\zeta}{R(\zeta)} & \ldots & \oint_{\ggh}\frac{\zeta^{N-1}f(\zeta)d\zeta}{R(\zeta)} &
\overline{\oint_{\ggh}\frac{f(\zeta)d\zeta}{R(\zeta)}} & \ldots & \overline{\oint_{\ggh}\frac{\zeta^{N-1}f(\zeta)d\zeta}{R(\zeta)}} &
\oint_{\ggh}\frac{f(\zeta)d\zeta}{(\zeta-\alpha_j)R(\zeta)}
\end{array}\right|
\end{equation}
and
\begin{equation}
\frac{\partial K(\alpha_j)}{\partial\mu}(\va,\mu,x,t)=
\frac{1}{2\pi i}\left|\begin{array}{ccccccc}
\oint_{\ggh_{m,1}}\frac{d\zeta}{R(\zeta)} & \ldots & \oint_{\ggh_{m,1}}\frac{\zeta^{N-1}d\zeta}{R(\zeta)} &
\overline{\oint_{\ggh_{m,1}}\frac{d\zeta}{R(\zeta)}} & \ldots & \overline{\oint_{\ggh_{m,1}}\frac{\zeta^{N-1}d\zeta}{R(\zeta)}} & \oint_{\ggh_{m,1}}\frac{d\zeta}{(\zeta-\alpha_j)R(\zeta)} \\
\ldots & \ldots & \ldots & \ldots & \ldots & \ldots & \ldots \\
\oint_{\ggh_{m,N}}\frac{d\zeta}{R(\zeta)} & \ldots & \oint_{\ggh_{m,N}}\frac{\zeta^{N-1}d\zeta}{R(\zeta)} &
\overline{\oint_{\ggh_{m,N}}\frac{d\zeta}{R(\zeta)}} & \ldots & \overline{\oint_{\ggh_{m,N}}\frac{\zeta^{N-1}d\zeta}{R(\zeta)}} &
\oint_{\ggh_{m,N}}\frac{d\zeta}{(\zeta-\alpha_j)R(\zeta)} \\
\oint_{\ggh_{c,1}}\frac{d\zeta}{R(\zeta)} & \ldots & \oint_{\ggh_{c,1}}\frac{\zeta^{N-1}d\zeta}{R(\zeta)} &
\overline{\oint_{\ggh_{c,1}}\frac{d\zeta}{R(\zeta)}} & \ldots & \overline{\oint_{\ggh_{c,1}}\frac{\zeta^{N-1}d\zeta}{R(\zeta)}} &
\oint_{\ggh_{c,1}}\frac{d\zeta}{(\zeta-\alpha_j)R(\zeta)} \\
\ldots & \ldots & \ldots & \ldots & \ldots & \ldots & \ldots \\
\oint_{\ggh_{c,N}}\frac{d\zeta}{R(\zeta)} & \ldots & \oint_{\ggh_{c,N}}\frac{\zeta^{N-1}d\zeta}{R(\zeta)} &
\overline{\oint_{\ggh_{c,N}}\frac{d\zeta}{R(\zeta)}} & \ldots & \overline{\oint_{\ggh_{c,N}}\frac{\zeta^{N-1}d\zeta}{R(\zeta)}} &
\oint_{\ggh_{c,N}}\frac{d\zeta}{(\zeta-\alpha_j)R(\zeta)} \\
\oint_{\ggh}\frac{f_{\mu}(\zeta)d\zeta}{R(\zeta)} & \ldots & \oint_{\ggh}\frac{\zeta^{N-1}f_{\mu}(\zeta)d\zeta}{R(\zeta)} &
\overline{\oint_{\ggh}\frac{f_{\mu}(\zeta)d\zeta}{R(\zeta)}} & \ldots & \overline{\oint_{\ggh}\frac{\zeta^{N-1}f_{\mu}(\zeta)d\zeta}{R(\zeta)}} &
\oint_{\ggh}\frac{f_{\mu}(\zeta)d\zeta}{(\zeta-\alpha_j)R(\zeta)}
\end{array}\right|,
\end{equation}
where $f_\mu$ is given by (\ref{eq_f'_mu}).

\begin{theorem} ($(\mu,x,t)$-Perturbation theorem)\\
Consider a simple contour $\gamma_0$ consisting of a finite union of oriented simple arcs $\gamma_0=\left(\bigcup\gg_{m,j}\right)\cup\left(\bigcup\gg_{c,j}\right)$
 with the distinct end points $\va_0$ and depending on parameters $\vb_0=(\mu_0,x_0,t_0)$ (see Figure \ref{fig:RHP_contours}). Assume $\va_0$ and $(\mu_0,x_0,t_0)$
satisfy a system of equations
\[
\vec{K}\left(\va_0,(\mu_0,x_0,t_0)\right)=\vec{0},
\label{eq:K_0}
\]
and $f$ is given by (\ref{eq:general f-function}).
Let $\gamma=\gamma(\va,\mu)$ be the contour of a RH problem which seeks a function
$h(z)$ which satisfies the following conditions
\begin{equation}
\left\{
\begin{array}{l}
h_{+}(z)+h_{-}(z)=2W_j,\ \mbox{on} \ \gg_{m,j}, \ \ j=0,1,...,N,\\
h_{+}(z)-h_{-}(z)=2\Omega_j,\ \mbox{on} \ \gg_{c,j}, \ \ j=1,...,N,\\
h(z)+f(z)\ \mbox{is analytic in}\ \mathbb{\overline{C}}\backslash\gg,
\end{array}
\right. \label{eq:RHP_NLS_h}
\end{equation}
where $\Omega_j=\Omega_j(\va,\mu,x,t)$ and $W_j=W_j(\va,\mu,x,t)$ are real constants whose numerical values will be determined from the RH conditions.
Assume that there is a function $h(z,\va_0,\mu_0,x_0,t_0)$ which satisfies (\ref{eq:RHP_h}) and suppose $\frac{h'(z,\va_0,\mu_0,x_0,t_0)}{R(z,\va_0)}\neq 0$
for all $z$ on $\gamma$.

Then the solution $\va=\va(\mu,x,t)$ of the system
\begin{equation}
\vec{K}\left(\va,(\mu,x,t)\right)=\vec{0}
\label{eq:vec_K_system}
\end{equation}
and $h(z,\va(\mu,x,t),\mu,x,t)$ which solves (\ref{eq:RHP_h}) are uniquely defined and smooth in $(\mu,x,t)$ in some
neighborhood of $(\mu_0,x_0,t_0)$.

Moreover, $\Omega_j(\mu,x,t)=\Omega_j(\va(\mu,x,t),\mu,x,t)$, and $W_j(\mu,x,t)=W_j(\va(\mu,x,t),\mu,x,t)$
are defined and smooth in $(\mu,x,t)$ in some neighborhood of $(\mu_0,x_0,t_0)$.

Furthermore, for $k\ge 1$:
\begin{equation}
\frac{\partial\alpha_j}{\partial\mu}(\mu,x,t)=-\frac{2\pi i \frac{\partial K(\alpha_j,\va,(\mu,x,t))}{\partial\mu}}{D(\va(\mu,x,t),\mu,x,t) \oint_{\ggh(\mu)}\frac{f'(\zeta)}{(\zeta-\alpha_j(\mu,x,t))R(\zeta,\va(\mu,x,t))}d\zeta},
\end{equation}
\begin{equation}
\frac{\partial h}{\partial\mu}(z,\va,\mu,x,t)=\frac{R(z,\va)}{2\pi i}\int_{\ggh(\mu)}
\frac{\frac{\partial f}{\partial\mu}(\xi,\mu,x,t)}{(\xi-z)R(\xi,\va)}d\xi,
\end{equation}
where $z$ is inside of $\ggh$,
\begin{equation}
\frac{\partial\Omega_j}{\partial\mu}(\mu,x,t)=\frac{-1}{D}
\left|\begin{array}{cccccc}
\oint_{\ggh_{m,1}}\frac{d\zeta}{R(\zeta)} & \ldots & \oint_{\ggh_{m,1}}\frac{\zeta^{N-1}d\zeta}{R(\zeta)} &
\overline{\oint_{\ggh_{m,1}}\frac{d\zeta}{R(\zeta)}} & \ldots & \overline{\oint_{\ggh_{m,1}}\frac{\zeta^{N-1}d\zeta}{R(\zeta)}} \\
\ldots & \ldots & \ldots & \ldots & \ldots & \ldots \\
\oint_{\ggh_{m,N}}\frac{d\zeta}{R(\zeta)} & \ldots & \oint_{\ggh_{m,N}}\frac{\zeta^{N-1}d\zeta}{R(\zeta)} &
\overline{\oint_{\ggh_{m,N}}\frac{d\zeta}{R(\zeta)}} & \ldots & \overline{\oint_{\ggh_{m,N}}\frac{\zeta^{N-1}d\zeta}{R(\zeta)}}  \\
\oint_{\ggh_{c,1}}\frac{d\zeta}{R(\zeta)} & \ldots & \oint_{\ggh_{c,1}}\frac{\zeta^{N-1}d\zeta}{R(\zeta)} &
\overline{\oint_{\ggh_{c,1}}\frac{d\zeta}{R(\zeta)}} & \ldots & \overline{\oint_{\ggh_{c,1}}\frac{\zeta^{N-1}d\zeta}{R(\zeta)}} \\
\ldots & \ldots & \ldots & \ldots & \ldots & \ldots  \\
\oint_{\ggh_{c,j-1}}\frac{d\zeta}{R(\zeta)} & \ldots & \oint_{\ggh_{c,j-1}}\frac{\zeta^{N-1}d\zeta}{R(\zeta)} &
\overline{\oint_{\ggh_{c,j-1}}\frac{d\zeta}{R(\zeta)}} & \ldots & \overline{\oint_{\ggh_{c,j-1}}\frac{\zeta^{N-1}d\zeta}{R(\zeta)}} \\
{\oint_{\ggh}\frac{ f_{\mu}(\xi)}{R(\xi,\va)}d\xi} & \ldots & {\oint_{\ggh}\frac{\xi^{N-1} f_{\mu}(\xi)}{R(\xi,\va)}d\xi}& \overline{\oint_{\ggh}\frac{ f_{\mu}(\xi)}{R(\xi,\va)}d\xi} & \ldots &
\overline{\oint_{\ggh}\frac{\xi^{N-1} f_{\mu}(\xi)}{R(\xi,\va)}d\xi}\\
\oint_{\ggh_{c,j+1}}\frac{d\zeta}{R(\zeta)} & \ldots & \oint_{\ggh_{c,j+1}}\frac{\zeta^{N-1}d\zeta}{R(\zeta)} &
\overline{\oint_{\ggh_{c,j+1}}\frac{d\zeta}{R(\zeta)}} & \ldots & \overline{\oint_{\ggh_{c,j+1}}\frac{\zeta^{N-1}d\zeta}{R(\zeta)}} \\
\ldots & \ldots & \ldots & \ldots & \ldots & \ldots  \\
\oint_{\ggh_{c,N}}\frac{d\zeta}{R(\zeta)} & \ldots & \oint_{\ggh_{c,N}}\frac{\zeta^{N-1}d\zeta}{R(\zeta)} &
\overline{\oint_{\ggh_{c,N}}\frac{d\zeta}{R(\zeta)}} & \ldots & \overline{\oint_{\ggh_{c,N}}\frac{\zeta^{N-1}d\zeta}{R(\zeta)}}
\end{array}\right|,
\label{eq:O_mu_thm}
\end{equation}
\begin{equation}
\frac{\partial W_j}{\partial\mu}(\mu,x,t)=\frac{-1}{D}
\left|\begin{array}{cccccc}
\oint_{\ggh_{m,1}}\frac{d\zeta}{R(\zeta)} & \ldots & \oint_{\ggh_{m,1}}\frac{\zeta^{N-1}d\zeta}{R(\zeta)} &
\overline{\oint_{\ggh_{m,1}}\frac{d\zeta}{R(\zeta)}} & \ldots & \overline{\oint_{\ggh_{m,1}}\frac{\zeta^{N-1}d\zeta}{R(\zeta)}} \\
\ldots & \ldots & \ldots & \ldots & \ldots & \ldots \\
\oint_{\ggh_{m,j-1}}\frac{d\zeta}{R(\zeta)} & \ldots & \oint_{\ggh_{m,j-1}}\frac{\zeta^{N-1}d\zeta}{R(\zeta)} &
\overline{\oint_{\ggh_{m,j-1}}\frac{d\zeta}{R(\zeta)}} & \ldots & \overline{\oint_{\ggh_{m,j-1}}\frac{\zeta^{N-1}d\zeta}{R(\zeta)}} \\
{\oint_{\ggh}\frac{ f_{\mu}(\xi)}{R(\xi,\va)}d\xi} & \ldots & {\oint_{\ggh}\frac{\xi^{N-1} f_{\mu}(\xi)}{R(\xi,\va)}d\xi}& \overline{\oint_{\ggh}\frac{ f_{\mu}(\xi)}{R(\xi,\va)}d\xi} & \ldots &
\overline{\oint_{\ggh}\frac{\xi^{N-1} f_{\mu}(\xi)}{R(\xi,\va)}d\xi}\\
\oint_{\ggh_{m,j+1}}\frac{d\zeta}{R(\zeta)} & \ldots & \oint_{\ggh_{m,j+1}}\frac{\zeta^{N-1}d\zeta}{R(\zeta)} &
\overline{\oint_{\ggh_{m,j+1}}\frac{d\zeta}{R(\zeta)}} & \ldots & \overline{\oint_{\ggh_{m,j+1}}\frac{\zeta^{N-1}d\zeta}{R(\zeta)}} \\
\ldots & \ldots & \ldots & \ldots & \ldots & \ldots  \\
\oint_{\ggh_{m,N}}\frac{d\zeta}{R(\zeta)} & \ldots & \oint_{\ggh_{m,N}}\frac{\zeta^{N-1}d\zeta}{R(\zeta)} &
\overline{\oint_{\ggh_{m,N}}\frac{d\zeta}{R(\zeta)}} & \ldots & \overline{\oint_{\ggh_{m,N}}\frac{\zeta^{N-1}d\zeta}{R(\zeta)}}  \\
\oint_{\ggh_{c,1}}\frac{d\zeta}{R(\zeta)} & \ldots & \oint_{\ggh_{c,1}}\frac{\zeta^{N-1}d\zeta}{R(\zeta)} &
\overline{\oint_{\ggh_{c,1}}\frac{d\zeta}{R(\zeta)}} & \ldots & \overline{\oint_{\ggh_{c,1}}\frac{\zeta^{N-1}d\zeta}{R(\zeta)}} \\
\ldots & \ldots & \ldots & \ldots & \ldots & \ldots  \\
\oint_{\ggh_{c,N}}\frac{d\zeta}{R(\zeta)} & \ldots & \oint_{\ggh_{c,N}}\frac{\zeta^{N-1}d\zeta}{R(\zeta)} &
\overline{\oint_{\ggh_{c,N}}\frac{d\zeta}{R(\zeta)}} & \ldots & \overline{\oint_{\ggh_{c,N}}\frac{\zeta^{N-1}d\zeta}{R(\zeta)}}
\end{array}\right|,
\label{eq:W_mu_thm}
\end{equation}
where $R(\xi)=R(\xi,\va)$.
\label{thm:perturb_NLS}
\end{theorem}

\begin{figure}
\begin{center}
\includegraphics[height=5cm]{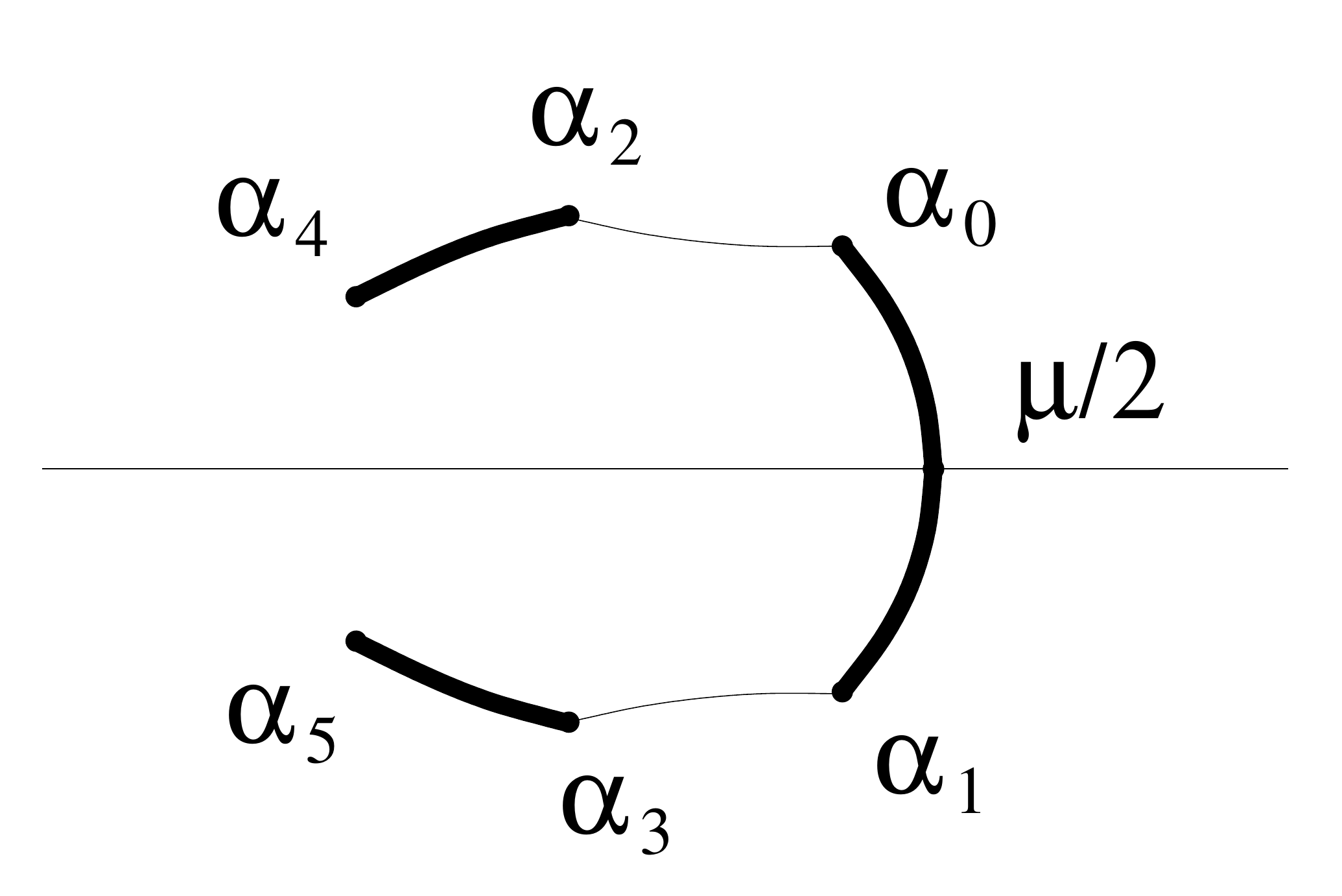}
\caption{\label{fig:NLS_genus_2_contour} The jump contour in the case of genus 2 with complex
conjugate symmetry in the notation of \cite{TVZzero}.}
\end{center}
\end{figure}

\begin{remark}
In the genus 2 case ($N=1$), taking into account the symmetry
\begin{equation}
\alpha_{2j}=\overline{\alpha}_{2j+1}, \ \ j=0,1,2,
\end{equation}
in the notation of \cite{TVZzero} where the numeration of $\alpha$'s was different,
$\ggh_m$ is a loop around the main arc $[\alpha_2,\alpha_4]$
and its complex conjugate $[\alpha_3,\alpha_5]$, and $\ggh_c$ is loop around the complementary arc $[\alpha_0,\alpha_2]$
and its complex conjugate $[\alpha_1,\alpha_3]$ (see Fig. \ref{fig:NLS_genus_2_contour}).

Then the equations (\ref{eq:vec_K_system}-\ref{eq:W_mu_thm}) read
\begin{equation}
\left|\begin{array}{ccc}
\oint_{\ggh_{m}}\frac{d\zeta}{R(\zeta)} & \oint_{\ggh_{m}}\frac{\zeta d\zeta}{R(\zeta)}& \oint_{\ggh_{m}}\frac{d\zeta}{(\zeta-\alpha_j)R(\zeta)} \\
\oint_{\ggh_{c}}\frac{d\zeta}{R(\zeta)} & \oint_{\ggh_{c}}\frac{\zeta d\zeta}{R(\zeta)}& \oint_{\ggh_{c}}\frac{d\zeta}{(\zeta-\alpha_j)R(\zeta)} \\
\oint_{\ggh(\mu)}\frac{f(\zeta)}{R(\zeta)}d\zeta & \oint_{\ggh(\mu)}\frac{\zeta f(\zeta) }{R(\zeta)}d\zeta & \oint_{\ggh(\mu)}\frac{f(\zeta)}{(\zeta-\alpha_j)R(\zeta)}d\zeta
\end{array}\right|=0,
\end{equation}
\begin{equation}
\frac{\partial\alpha_j}{\partial\mu}(\mu,x,t)=-\frac{2\pi i}{D}\frac{\left|\begin{array}{ccc}
\oint_{\ggh_{m}}\frac{d\zeta}{R(\zeta)} & \oint_{\ggh_{m}}\frac{\zeta d\zeta}{R(\zeta)}& \oint_{\ggh_{m}}\frac{d\zeta}{(\zeta-\alpha_j)R(\zeta)} \\
\oint_{\ggh_{c}}\frac{d\zeta}{R(\zeta)} & \oint_{\ggh_{c}}\frac{\zeta d\zeta}{R(\zeta)}& \oint_{\ggh_{c}}\frac{d\zeta}{(\zeta-\alpha_j)R(\zeta)} \\
\oint_{\ggh(\mu)}\frac{\frac{\partial f}{\partial\mu}(\zeta)}{R(\zeta)}d\zeta & \oint_{\ggh(\mu)}\frac{\zeta \frac{\partial f}{\partial\mu}(\zeta) }{R(\zeta)}d\zeta & \oint_{\ggh(\mu)}\frac{\frac{\partial f}{\partial\mu}(\zeta)}{(\zeta-\alpha_j)R(\zeta)}d\zeta
\end{array}\right|}{\oint_{\ggh(\mu)}\frac{f'(\zeta)}{(\zeta-\alpha_j)R(\zeta)}d\zeta},
\end{equation}
\begin{equation}
\frac{\partial h}{\partial\mu}(z,\va,\mu,x,t)=\frac{R(z,\va)}{2\pi i}\int_{\ggh(\mu)}
\frac{\frac{\partial f}{\partial\mu}(\xi,\mu,x,t)}{(\xi-z)R(\xi,\va)}d\xi,
\end{equation}
where $z$ is inside of $\ggh$,
\begin{equation}
\frac{\partial\Omega}{\partial\mu}(\mu,x,t)=
-\frac{1}{D}
\left|\begin{array}{cc}
\oint_{\ggh_{m}}\frac{d\zeta}{R(\zeta)} & \oint_{\ggh_{m}}\frac{\zeta d\zeta}{R(\zeta)} \\
{\oint_{\ggh}\frac{ f_{\mu}(\xi)}{R(\xi,\va)}d\xi} & {\oint_{\ggh}\frac{\xi  f_{\mu}(\xi)}{R(\xi,\va)}d\xi}
\end{array}\right|,
\end{equation}
\begin{equation}
\frac{\partial W}{\partial\mu}(\mu,x,t)=
-\frac{1}{D}
\left|\begin{array}{cc}
{\oint_{\ggh}\frac{ f_{\mu}(\xi)}{R(\xi,\va)}d\xi} & {\oint_{\ggh}\frac{\xi  f_{\mu}(\xi)}{R(\xi,\va)}d\xi}\\
\oint_{\ggh_{c}}\frac{d\zeta}{R(\zeta)} & \oint_{\ggh_{c}}\frac{\zeta d\zeta}{R(\zeta)}
\end{array}\right|,
\end{equation}
\end{remark}
where $\alpha_j=\alpha_j(\mu,x,t)$, $R(\zeta)=R(\zeta,\va(\mu,x,t))$, $f(\zeta)=f(\zeta,\mu,x,t)$, $\frac{\partial f}{\partial\mu}(\zeta)=\frac{\partial f}{\partial\mu}(\zeta,\mu)$ and
\begin{equation}
D=D(\mu,x,t)=\left|\begin{array}{cc}
\oint_{\ggh_{m}}\frac{d\zeta}{R(\zeta)} & \oint_{\ggh_{m}}\frac{\zeta d\zeta}{R(\zeta)} \\
\oint_{\ggh_{c}}\frac{d\zeta}{R(\zeta)} & \oint_{\ggh_{c}}\frac{\zeta d\zeta}{R(\zeta)}
\end{array}\right|.
\end{equation}

\begin{remark}
The perturbation theorem \ref{thm:perturb_NLS} guarantees that the solution of the RHP (\ref{eq:RHP_NLS_h})
is uniquely continued with respect to external parameters. Additional sign conditions on $\Im h$ need to be
satisfied, for $h$ to correspond to an asymptotic solution of NLS as in \cite{TVZzero}. The sign conditions
have to be satisfied near $\gamma$ and additionally on a
semiinfinite complementary arcs connecting the end points of $\gamma$ to $\infty$.
\end{remark}

\section{Appendix}

\subsection{Alternative formulation of the scalar RHP: function $g'$}

An alternative approach is to start with a scalar RHP on $g'$ which we
formulate as the following: given an ordered sequence of distinct points $\va=\left(\alpha_0,...,\alpha_{2N+1}\right)$
and oriented arcs $\gamma_{m,j}$, $\gamma_{c,j}$ connecting them, find a function $g'(z)$ so
the following Riemann-Hilbert conditions are satisfied
\begin{equation}
\left\{
\begin{array}{l}
g'_{+}(z)+g'_{-}(z)=f'(z),\ \mbox{on} \ \gg_{m,j},\ \ j=0,1,...,N, \\
g'(z)\ \mbox{is analytic in}\ \mathbb{{C}}\backslash\bigcup\gg_{m,j},\\
g'(z)\sim O(z^{-2}),\ \mbox{as}\ z\to\infty,\\
\end{array}
\right. \label{eq:RHP_g'}
\end{equation}
where by $g'_\pm$ are denoted the limiting values of $g'$ approaching the contour from the positive and the negative sides
 respectively.

If $g(z)$ satisfies (\ref{eq:RHP_g'}) then $g(z)$ satisfies the following Riemann-Hilbert conditions
\begin{equation}
\left\{
\begin{array}{l}
g_{+}(z)+g_{-}(z)=f(z)+W_j,\ \mbox{on} \ \gg_{m,j},\ \ j=0,1,...,N,\\
g_{+}(z)-g_{-}(z)=\Omega_j,\ \mbox{on} \ \gg_{c,j},\ \ j=0,1,...,N,\\
g(z)\ \mbox{is analytic in}\ \mathbb{\overline{C}}\backslash\gg,\\
\end{array}
\right. \label{eq:RHP_g}
\end{equation}
where $\gamma=\left(\bigcup\gg_{m,j}\right)\cup\left(\bigcup\gg_{c,j}\right)$
is an oriented simple curve called the jump contour.
The constants $W_j$, $\Omega_j$ are found from the condition of $g(z)$ being analytic at $\infty$.

The conditions (\ref{eq:RHP_g'}) arguably are more natural since they do not involve unknown constants
$W_j$ and $\Omega_j$ which are found from the behavior at infinity.

\subsection{Evaluation of a simplified integral for Lemma \ref{lem:par_int_zero}}

Let $\beta_1=\mu$, and all other parameters $\beta_2,\ldots$ are constants and
\[
f(z,\mu)=c(\mu)(z-z_0)\log(z-z_0),
\]
where $z_0=z_0(\mu)$, $c(\mu)$ are continuously differentiable in $\mu$ and some branch
of the logarithm is chosen with the branchcut passing from $z_0$ to infinity and not
intersecting the contour $[z_1,z_0(\mu))\cup(z_0(\mu),z_2]$.

Consider the contour integral through $z_0$ with some fixed $z_1$ and $z_2$
\[
I_1(\mu)=\int_{[z_1,z_0(\mu)]\cup[z_0(\mu),z_2]}f(\xi,\mu)d\xi
\]
\[
=c(\mu)\int_{[z_1,z_0]\cup[z_0,z_2]} (\xi-z_0)\log(\xi-z_0) d\xi
\]
and after the change of variables $y=\xi-z_0$
\[
=c(\mu)\int_{[z_1-z_0,0]\cup[0,z_1-z_0]}y\log y \ dy.
\]

Thus
\[
I_1(\mu)=c(\mu)\left[\left(\log (z_2-z_0)-\frac{1}{2}\right)\frac{(z_2-z_0)^2}{2}
-\left(\log (z_1-z_0)-\frac{1}{2}\right)\frac{(z_1-z_0)^2}{2}
\right].
\]
The derivative of $I_1(\mu)$ in $\mu$ is
\[
\frac{d I_1}{d \mu}(\mu)
=\frac{c'(\mu)}{c(\mu)}I_1
+c(\mu)\left[
\left(\log (z_2-z_0)-\frac{1}{2}\right)(z_2-z_0)(-z'_0)
-\left(\log (z_1-z_0)-\frac{1}{2}\right)(z_1-z_0)(-z'_0)
\right].
\]
On the other hand
\[
\frac{\partial f}{\partial\mu}(z,\mu)=c'(\mu)\frac{f(z,\mu)}{c(\mu)}+c(\mu)\left[\log(z-z_0)(-z'_0)+(-z'_0)\right]
\]
and
\[
\int_{[z_1,z_0]\cup[z_0,z_2]}\frac{\partial f(\xi,\mu)}{\partial\mu} d\xi
= \frac{c'(\mu)}{c(\mu)}I_1(\mu) - (z_2-z_1)c(\mu)z'_0(\mu) -c(\mu)z'_0 \int_{[z_1-z_0,0]\cup[0,z_2-z_0]} \log y dy,
\]
where $y=\xi-z_0$
\[
= \frac{c'(\mu)}{c(\mu)}I_1(\mu) -c(\mu)z'_0 \left[ (z_2-z_0)\log (z_2-z_0) - (z_1-z_0)\log(z_1-z_0) \right]
= \frac{\partial I_1}{\partial \mu}.
\]

So
\[
\frac{d I_1}{d \mu}(\mu)=\int_{[z_1,z_0]\cup[z_0,z_2]}\frac{\partial f(\xi,\mu)}{\partial\mu} d\xi
\]
the derivative and the integral can be interchanged.

\end{document}